\documentclass[fleqn,usenatbib]{mnras}

\usepackage{newtxtext,newtxmath}

\usepackage[T1]{fontenc}

\DeclareRobustCommand{\VAN}[3]{#2}
\let\VANthebibliography\thebibliography
\def\thebibliography{\DeclareRobustCommand{\VAN}[3]{##3}\VANthebibliography}

\usepackage{graphicx}	
\usepackage{amsmath}	
\usepackage{relsize}
\usepackage{multirow}

\usepackage[normalem]{ulem}
\usepackage{xcolor}
\usepackage{soul}
\setstcolor{red}

\title[Flare State Black Hole Signatures]{Millimeter Observational Signatures of Flares in Magnetically Arrested Black Hole Accretion Models}

\usepackage{CJK}
\author[H. Jia et al.]{He Jia (\begin{CJK*}{UTF8}{gbsn}贾赫\end{CJK*})$^{1}$\thanks{E-mail: hejia@princeton.edu},
Bart Ripperda$^{2,3,4,1}$,
Eliot Quataert$^{1}$,
Christopher J. White$^{4,1}$,
Koushik Chatterjee$^{5,6}$,
\newauthor
Alexander Philippov$^{7}$
and Matthew Liska$^{6,8}$
\\
$^{1}$Department of Astrophysical Sciences, Princeton University, Princeton, NJ 08544, USA\\
$^{2}$School of Natural Sciences, Institute for Advanced Study, 1 Einstein Drive, Princeton, NJ 08540, USA\\
$^{3}$NASA Hubble Fellowship Program, Einstein Fellow\\
$^{4}$Center for Computational Astrophysics, Flatiron Institute, 162 Fifth Avenue, New York, NY 10010, USA\\
$^{5}$Black Hole Initiative, Harvard University, 20 Garden Street, Cambridge, MA 02138, USA\\
$^{6}$Center for Astrophysics, Harvard \& Smithsonian, 60 Garden Street, Cambridge, MA 02138, USA\\
$^{7}$Department of Physics, University of Maryland, College Park, MD 20742, USA\\
$^{8}$Institute for Theory and Computation, Harvard University, 60 Garden Street, Cambridge, MA 02138, USA
}

\date{Accepted XXX. Received YYY; in original form ZZZ}

\pubyear{2015}

\begin{document}
\label{firstpage}
\pagerange{\pageref{firstpage}--\pageref{lastpage}}
\maketitle

\begin{abstract}
In general relativistic magneto-hydrodynamic (GRMHD) simulations, accreted magnetic flux on the black hole horizon episodically decays, during which magnetic reconnection heats up the plasma near the horizon, potentially powering high-energy flares like those observed in M87* and Sgr A*.   We study the mm observational counterparts of such flaring episodes in very high-resolution GRMHD simulations.    The change in 230 GHz flux during the expected high energy flares depends primarily on the efficiency of accelerating 
$\gamma \gtrsim 100$ ($T_e \gtrsim 10^{11}$ K) electrons. For models in which the electrons are heated to $T_e \sim 10^{11}$ K during flares, 
the hot plasma produced by reconnection significantly enhances 230 GHz emission and increases the size of the 230 GHz image. By contrast, for models in which the electrons are heated to higher temperatures (which we argue are better motivated), the reconnection-heated plasma is too hot to produce significant 230 GHz synchrotron emission, and the 230 GHz flux decreases during high energy flares.   We do not find a significant change in the mm polarization during flares as long as the emission is Faraday thin.  We also present expectations for the ring-shaped image as observed by the Event Horizon Telescope during flares, as well as multi-wavelength synchrotron spectra.  Our results highlight several limitations of standard post-processing prescriptions for the electron temperature in GRMHD simulations.   We also discuss the implications of our results for current and future observations of flares in Sgr A*, M87*, and related systems.  Appendices contain detailed convergence studies with respect to resolution and plasma magnetization.

\end{abstract}




\section{Introduction}

Black holes are
often surrounded by accretion disks with relativistic jets emitting at a range of wavelengths from radio to $\gamma$-ray \citep[e.g.][]{narayan2005black,yuan2014hot,Davis2020Magnetohydrodynamics}.
In addition to quasi-steady emission, bright X-ray and $\gamma$-ray flares \citep[e.g.][]{harris2011experiment,Abramowski2012very} are observed from Low Luminosity Active Galactic Nuclei such as M87*.  Sgr A* exhibits analogous flaring in the infrared (IR) and X-ray \citep[e.g.][]{Yusef-Zadeh2009Simultaneous,Yusef-Zadeh2010Occultation,Trap2011Concurrent,fazio2018multiwavelength}.

The mechanism of such high energy flares is not fully understood.
In magnetically arrested disk (MAD) models \citep{Igumenshchev2003Three-dimensional,Narayan2003Magnetically,Tchekhovskoy2011Efficient} episodic dissipation of magnetic energy near the horizon is a key dynamical feature of the accretion flow:  magnetic flux and magnetic energy build up on the black hole horizon until they become strong enough to suppress accretion.   Instabilities (e.g., magnetic Rayleigh-Taylor) and reconnection then set in episodically (in ``flux eruptions''), regulating the amount of magnetic flux and energy stored near the black hole.   The electromagnetic energy released through this reconnection is a promising source of  observed flares from black holes \citep{Dodds-Eden2010Time-Dependent,Dexter2020Sgr,Chatterjee:2021,Porth2021Flares,Chatterjee:2022,ripperda2022black,hakobyan2022radiative,Scepi:2022}.

Using Very Long Baseline Interferometry (VLBI) observations at 230 GHz, the Event Horizon Telescope (EHT) Collaboration presented the first images of the plasma around the supermassive black holes in M87* \citep{eht2019m87i} and Sgr A* \citep{eht2022sgri}.
For M87, the polarization maps have also been released \citep{eht2021m87vii}.
For M87* in particular the observations generally favor MAD models. For Sgr A*, the observational situation is less clear as no model is consistent with all the observations \citep{eht2022sgrv}, but theoretical models of the fueling of Sgr A$^{*}$ by stellar winds predict that the flow becomes magnetically arrested in the inner accretion region (\citealt{Ressler_2020}).  Numerical models also suggest that the episodic magnetic flux eruptions in MAD models can explain many aspects of the episodic infrared and X-ray flares observed in Sgr A*. In particular,   \citet{Dexter2020Sgr} and \citet{Porth2021Flares} showed that such models can qualitatively explain the motion of the IR center-of-light and rotation in the linear polarization direction seen by the VLT interferometer GRAVITY during IR flares from Sgr A* \citep{Gravity2018}.


It is not clear how horizon-scale observables accessible to EHT will change during the {magnetic} flux eruptions characteristic of MAD models.  If the magnetic flux eruptions indeed drive high-energy flares in Sgr A*, M87*, and other systems, connecting the mm observables to higher energy observables will be a key test of theoretical models.  In this paper, we aim to bridge this gap
and study multi-wavelength observational signatures
of flux eruptions, with a focus on the relation between 230 GHz EHT observables and higher energy radiation.   Throughout this paper we will refer to the flux eruptions interchangeably as ``flares" by which we specifically mean high-energy flares.   We explain our motivation for this identification in more detail in Section \ref{sec:lightcurve} but we also stress that this identification has not yet been conclusively established and more work directly predicting the high energy radiation from simulations is necessary to do so.

The remainder of this paper is organized as follows.
The methodology and numerical techniques are presented in Section \ref{sec:method}.
We present 230 GHz light curves in Section \ref{sec:lightcurve}, 230 GHz polarized images in Section \ref{sec:image}, and sychrotron emission spectra in Section \ref{sec:multi}.   We conclude 
in Section \ref{sec:discuss} with a discussion on the appearance of flux eruptions at millimeter wavelengths, under which conditions the millimeter emission brightens or dims during high-energy flares, and how modeling the emission can be further improved.
The Appendices contain detailed convergence studies with respect to resolution and plasma magnetization (see \S \ref{sec:method} for a brief summary).

\section{Methodology}

\label{sec:method}

Recently, \citet{ripperda2022black} conducted high resolution (dubbed \textit{extreme} resolution in their paper) general relativistic magneto-hydrodynamic (GRMHD) simulations, which for the first time captured plasmoid-mediated reconnection in a 3D magnetically arrested disk, during the episodic magnetic flux eruptions.  The simulations employ Static Mesh Refinement (SMR) for spherical Kerr-Schild coordinates $r,\theta,\phi$ describing a Kerr black hole with dimensionless spin $a=0.9375$ on a numerical grid with resolution $N_r \times N_\theta \times N_\phi =  5376\times2304\times2304$. The radial domain is fixed to $[1.2, 2000]\,r_g$. The GRMHD equations are integrated until $10000\,r_{\rm g}/c$. 
A ceiling is enforced to maintain $\sigma \leq \sigma_{\rm floor}=25$, where the magnetization $\sigma$ is defined using the magnetic field strength $b$ co-moving with the fluid, and fluid-frame rest-mass density $\rho$,
\begin{equation}
    \sigma\equiv b^2/(4 \pi \rho c^2).
\end{equation}
A pure ionized hydrogen composition is assumed, and the equation of state is that of an ideal gas with an adiabatic index of $\hat{\gamma} = 13/9$. The simulation is initialized to reach a MAD state, showing large periods of accretion where magnetic flux piles up on the horizon and quasi-periodic short flux eruptions where magnetic energy dissipates through magnetic reconnection. This dissipated magnetic energy can heat the plasma and potentially power multiwavelength flares.

GRMHD simulations do not predict the electron temperature which is required for calculating  synchrotron emission.   We use the following $R_{\rm high}-R_{\rm low}$ model motivated by phenomenological considerations \citep{moscibrodzka2016general} to compute the electron temperature from GRMHD fluid pressure $p$, density $\rho$ and plasma $\beta$,
\begin{align}
    T_e = \frac{2\,T_{\rm fluid}}{1+R},\ \ {\rm where}\ \ T_{\rm fluid} &\equiv \frac{m_p p}{2\,k_B\rho}, \nonumber \\
    R &\equiv \frac{T_p}{T_e} = \frac{\beta^2}{1+\beta^2} R_{\rm high} + \frac{1}{1+\beta^2} R_{\rm low}, \nonumber \\
    \beta &\equiv \frac{8\pi p}{B^2}.
    \label{eq:R}
\end{align}
Note that larger $R$ models have smaller $T_e/T_{\rm fluid}$, and vice versa.    In our modelling we assume fixed $R_{\rm high}$ and $R_{\rm low}$, although in reality the relation between $T_e$ and $T_{\rm fluid}$ is more complicated and could well be time and/or space-dependent.   Indeed, we shall see that our analysis of the simulated mm variability {during a magnetic flux eruption} highlights that it is sensitive to the possibility of temporal and/or spatial variability of $T_e/T_{\rm fluid}$.   This implies that standard $R_{\rm high}-R_{\rm low}$ post-processing prescriptions are limited in their ability to predict the variability associated with the distinctive magnetic flux eruptions present in MAD models.  

Our calculations only include synchrotron emission from thermal electrons;  while this is likely reasonable for the 230 GHz modelling in Sections \ref{sec:lightcurve}-\ref{sec:image}, at higher frequencies non-thermal electrons and inverse Compton emission become more important, so our spectral modelling results in Section \ref{sec:multi} likely represents lower limits to higher frequency emission instead of quantitatively precise predictions.

We generate ray tracing images from the GRMHD data with the \texttt{blacklight} code \citep{white2022blacklight}, which integrates the radiation transfer equations along geodesics to obtain the observed intensity and polarization maps.\footnote{We adopt the fast light approximation which assumes that the speed of light is infinite. This should be fine for our purposes as we mainly study the evolution of emission on the timescale of $\mathcal{O}(10^2)\,M$.}
We ignore the plasma outside $105\,r_g$ where the emission is negligible at the wavelengths studied in this paper.
The spatial resolution of the GRMHD data is reduced by a factor of $4\times 4\times 4$, i.e. only one of four successive points along each spatial dimension is kept, to speed up ray tracing computation.
We also ignore the region with $\sigma >\sigma_{\rm cut}$ where the temperature is not reliable since the plasma may be governed by the advection of injected density and pressure due to the sigma ceiling ($\sigma_{\rm floor}$=25); we choose $\sigma_{\rm cut}\!=\!1$ in Sections \ref{sec:lightcurve}-\ref{sec:image}, and $\sigma_{\rm cut}\!=\!10$ in Section \ref{sec:multi} since at higher frequencies the emission may be dominated by $\sigma\sim\sigma_{\rm cut}$ regions.\footnote{While both $\sigma_{\rm floor}$ and $\sigma_{\rm cut}$ represent a {ceiling} for the plasma magnetization $\sigma$, in this paper $\sigma_{\rm floor}$ stands for the numerical floor applied in the GRMHD simulation, while $\sigma_{\rm cut}$ represents the cutoff applied during ray tracing computation. {Note that in reality, $\sigma$ in the magnetospheric and jet regions is likely much higher than the value $\sigma_{\rm floor}$ used in GRMHD simulations.}}    We explore the convergence of our results with respect to the choice of GRMHD resolution and $\sigma_{\rm cut}$ in Appendix \ref{sec:converge}.  The convergence with respect to both GRMHD resolution and $\sigma_{\rm cut}$ depends on both the frequency of the radiation and the assumed mapping between electron temperature and GRMHD fluid temperature, but standard resolution ($\sim256^3$) GRMHD simulations and $\sigma_{\rm cut}\sim1$ are likely adequate for most of current EHT analyses.    Models with $R \lesssim 100$ are well-converged at 230 GHz while models with $R \simeq 100$ show some weak dependence on both resolution and $\sigma_{\rm cut}$.  All models show some dependence on $\sigma_{\rm cut}$ for higher energy radiation because the high $\sigma$ regions tend to have high temperatures in our models, which mostly emit higher frequency synchrotron radiation.

\begin{figure*}
	\includegraphics[width=15cm]{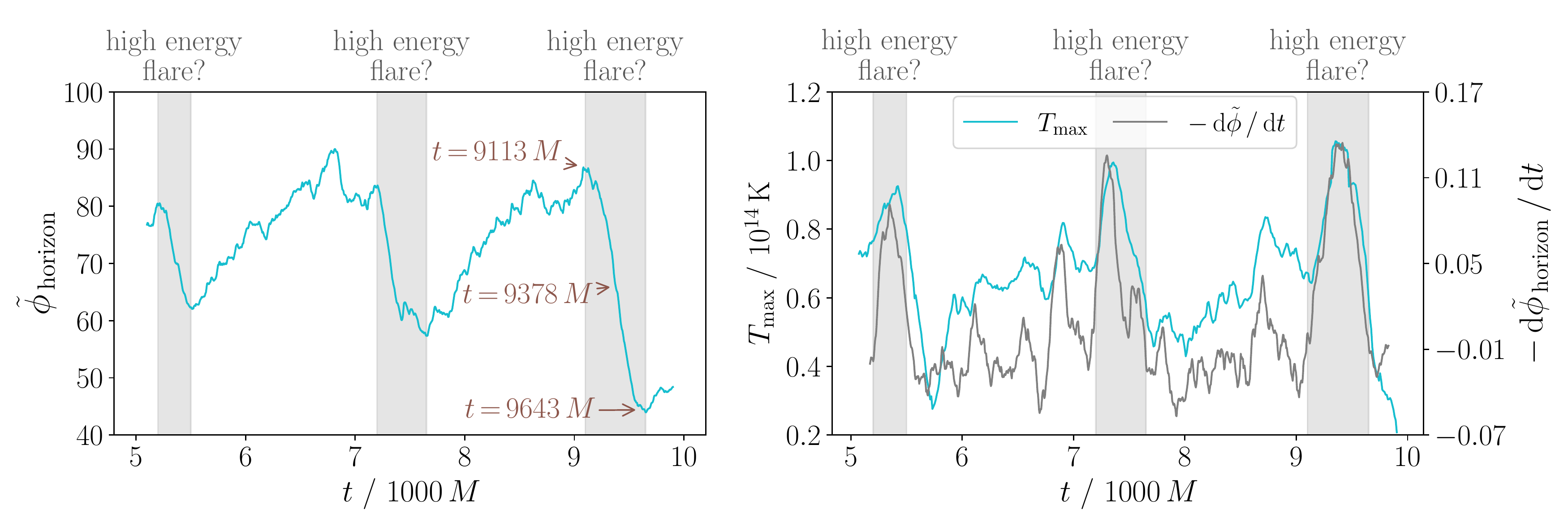}
    \caption{GRMHD fluid properties as a function of time, smoothed by a moving average with a window of 150$\,M$. {\em Left:} the magnetic flux on the black hole horizon $\tilde{\phi}_{\rm horizon} \equiv \frac{1}{2} \int_0^{2\pi} \int_0^{\pi} \left| ^*F^{rt}\right|\sqrt{-g}{\rm d}\theta{\rm d}\phi$.
    {\em Right:} the maximum fluid temperature over the whole GRMHD grid, which is used as a proxy for the amount of heated/accelerated plasma.
    The grey bands indicate the three major magnetic-flux decay states.
    We use the correlation between $T_{\max}$ and $-\,{\rm d}\tilde{\phi}_{\rm \,horizon}\,/\,{\rm d}t$ during flux eruptions as a proxy for the timing of high energy flares in systems like M87* and Sgr A*: the energy released by reconnection heats up the plasma during the magnetic flux decay.
   Note however, that in reality, $\sigma \gg \sigma_{\rm floor}$ in the jet and the temperature increase due to flux eruptions may be much larger than shown here (and regulated by strong radiative cooling). We also find a similar trend of increasing temperature during magnetic flux decay for the 90\% and 99\% percentiles of $T_{\rm fluid}$.}
    \label{fig:fluid}
\end{figure*}

We choose M87* parameters for ray tracing, with black hole mass $M=6.5 \times 10^9\,M_{\odot}$ and distance $D=16.8\,\mathrm{Mpc}$ \citep{blakeslee2009acs,bird2010inner,cantiello2018next}, since the spatially resolved intensity and polarization in M87* are better constrained by EHT data than Sgr A*.   However, our results on the evolution of the 230 GHz emission during flux eruptions are generic and apply to Sgr A* as well.   We will discuss  the application to Sgr A* in more detail in Section \ref{sec:discuss}.
In our ray tracing calculations the camera is located at $r_0=100\,r_g$, $\theta_0=163^{\circ}$ \citep{mertens2016kinematics} and $\phi_0=0^{\circ}$, while the approaching jet has a position angle of $288^{\circ}$ \citep{walker2018structure}, pointing towards the right and slightly up in the images. {Since the disk structure of a MAD during a flux eruption is highly non-axisymmetric, the corresponding raytraced image depends on the azimuthal position of the camera. However, this dependence is relatively weak for the low inclination case of M87* (see \citealt{Gelles:2022} for more details), and therefore, our results for $\phi_0=0^{\circ}$ should hold for other $\phi_0$.}
Since (ideal) GRMHD simulations are dimensionless, the normalization factor between code units and physical units needs to be set using the observed flux of emission.
Unless otherwise specified, the overall density in the simulation is normalized such that the averaged 230 GHz flux equals 0.66 Jy \citep{eht2019m87iv}.
All the plots in this paper showing the evolution with time are smoothed by a moving average with a window of 150$\,M$ (15 snapshots), so that the general trends are more clearly presented.

In order to quantify the variability of the predicted images between quiescent and flare states, we compute the following statistics for the 230 GHz images blurred with a 20 $\mu$as Gaussian kernel, similar to those used in EHT analysis \citep{chael2018interferometric,eht2019m87iv,eht2021m87viii}; see Section 2 of \cite{jia2022observational} for more details about how these quantities are measured from the images.
\begin{enumerate}
    \item The ring diameter $d$, determined by the average of peak intensity along different directions of the ring.
    \item The ring width $w$, defined as the Full Width Half Maximum (FWHM) of the intensity map, averaged over different directions.
    \item The ring orientation $\eta$ and degree of asymmetry $A$,
    \begin{align}
        \eta=\left< \mathrm{Arg} \left[ \int_0^{2\pi} I(\theta) e^{i\theta} d\theta \right] \right>_{r \in [r_{\mathrm{in}}, r_{\mathrm{out}}]},\
        \label{eq:eta}
    \end{align}
    \begin{equation}
        A=\left< \frac{ \left|\, \int_0^{2\pi} I(\theta) e^{i\theta}d\theta \,\right| }{\int_0^{2\pi} I(\theta) d\theta} \right>_{r \in [r_{\mathrm{in}}, r_{\mathrm{out}}]},
        \label{eq:A}
    \end{equation}
    where $r_{\rm in}$ and $r_{\rm out}$ are the radii where the intensity drops to half of the peak value along that direction.
    \item The fractional central brightness $f_C$,
    \begin{equation}
        f_C=\frac{\left< I(r,\theta) \right>_{\theta \in \left[ 0, 2\pi \right],\, r\in \left[ 0, 5\mu\mathrm{as} \right] }}{ \left< I(d/2, \theta) \right>_{\theta \in \left[ 0, 2\pi \right]} }.
        \label{eq:f_C}
    \end{equation}
    \item The pixel-level image-averaged linear polarization fraction,
    \begin{equation}
        \left<|m|\right>=\frac{\mathlarger{\sum} _i \sqrt{\mathcal{Q}^2_i + \mathcal{U}^2_i}}{\mathlarger{\sum} _i \mathcal{I}_i},
        \label{eq:m}
    \end{equation}
    where the Stokes $\mathcal{I}$, $\mathcal{Q}$ and $\mathcal{U}$ are summed over all the pixels and snapshots.
    \item The $\beta_m$ polarization statistics \citep{Palumbo2020Discriminating} defined in polar coordinates $(\rho,\,\phi)$ of the image plane,
    \begin{equation}
        \beta_m=\frac{1}{I_{\mathrm{ann}}} \int_{\rho_{\min}}^{\rho_{\max}} \int_0^{2\pi} (\mathcal{Q} + i\mathcal{U}) e^{-im\phi} \rho d\phi d\rho,
        \label{eq:beta_m}
    \end{equation}
    where we take $m=2$, $\rho_{\min}=0$, $\rho_{\max}\to\infty$, and $I_{\mathrm{ann}}$ is the total intensity flux between $\rho_{\min}$ and $\rho_{\max}$.
    Note that $\beta_2$ quantifies the orientation of the polarization and is widely used to constrain the magnetic field structure around the black hole (see Equations \ref{eq:etaB}-\ref{eq:argbeta2} and the discussions therein).
\end{enumerate}

\begin{figure*}
	\includegraphics[width=16cm]{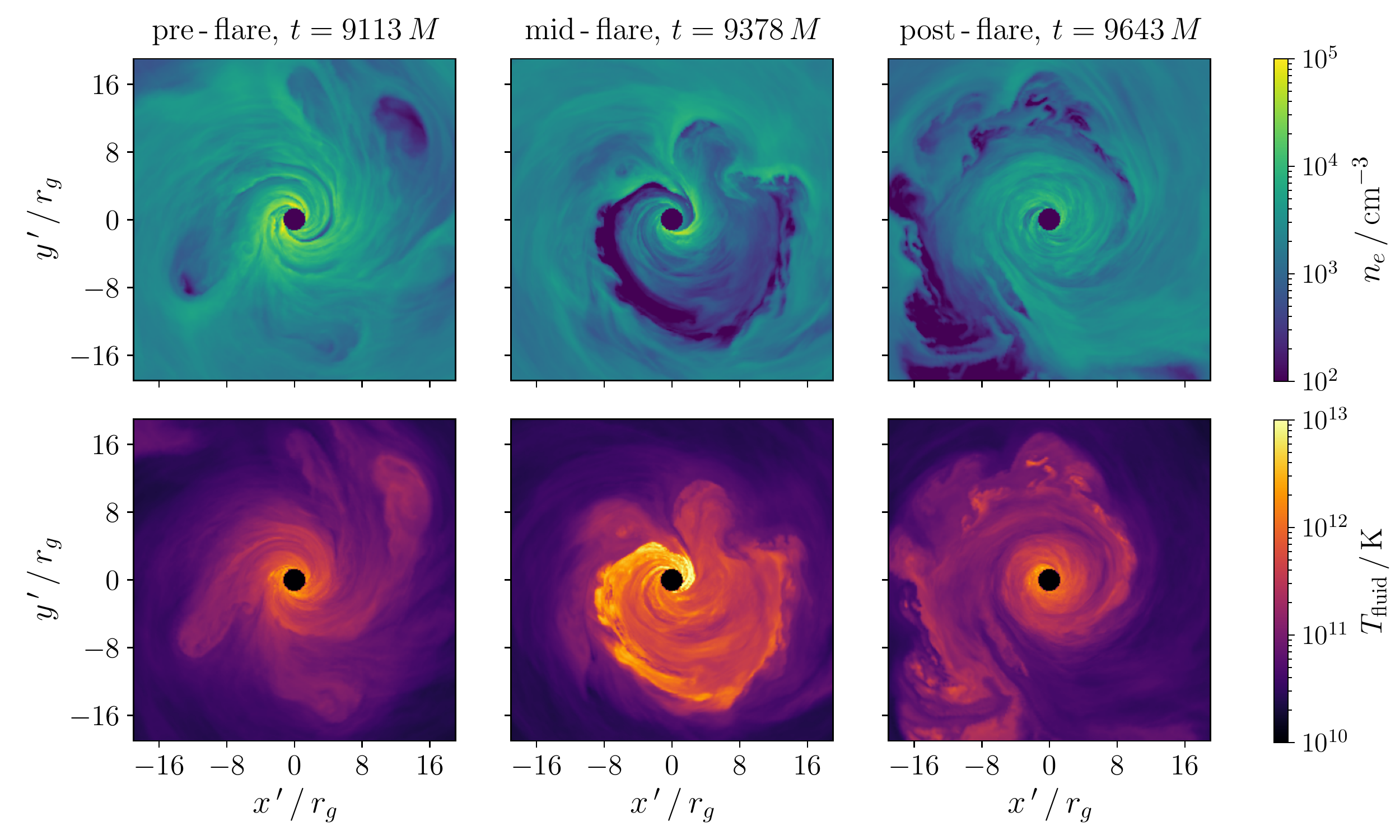}
    \caption{Equatorial slices of density and temperature in the GRMHD simulation for the pre-flare, mid-flare and post-flare snapshots.
    The $x'-y'$ coordinates are rotated to match the ray tracing images in Figure \ref{fig:polar-07-00}.
    We average over the fluid between $\pm 15^{\circ}$ from the midplane to capture the structures that are not exactly on the midplane, while for $T_{\rm fluid}$ the average is weighted by $n_e$.
    The normalization of $n_e$ is set based on the 230 GHz flux of $R=1$ model, which is $4.88\,(156)$ times larger if we use $R=10\,(100)$ instead of $R=1$.}
    \label{fig:nT}
\end{figure*}

\begin{figure*}
	\includegraphics[width=15cm]{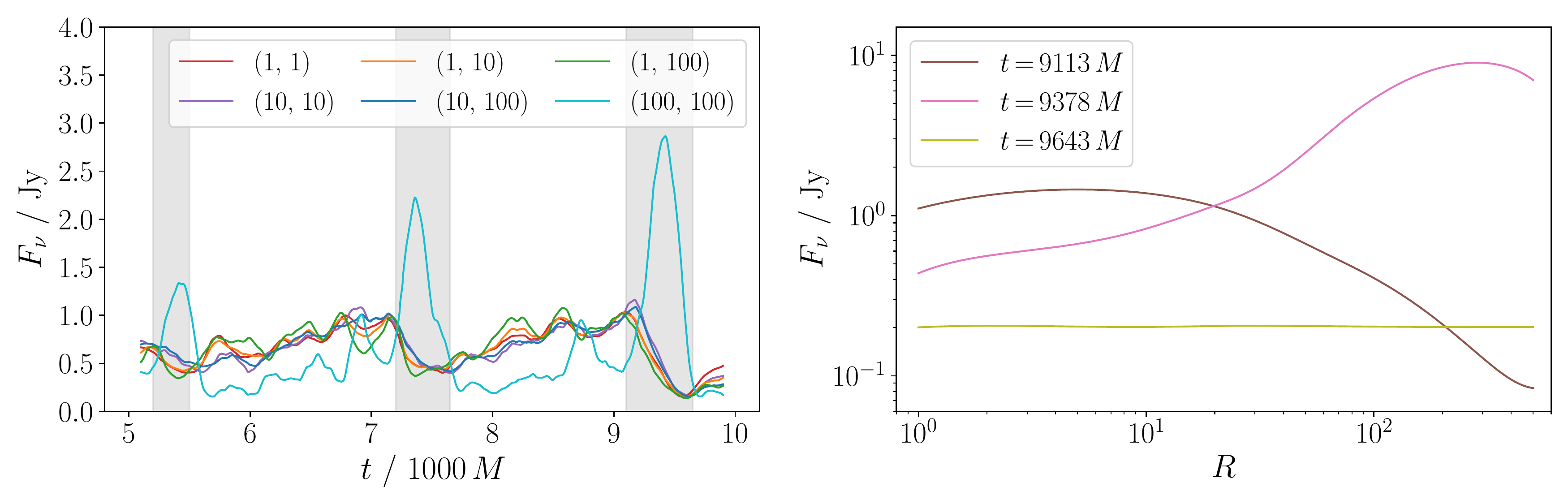}
    \caption{The evolution of 230 GHz flux with different electron temperature models.
    {\em Left:} the 230 GHz flux as a function of time, smoothed by a moving average with a window of 150$\,M$, with the legend indicating ($R_{\rm low}, R_{\rm high}$).
    {\em Right:} the 230 GHz flux at the beginning, midpoint and end of a flux decay state, as a function of $R=R_{\rm low}=R_{\rm high}$.
    Here we adjust the density normalization for each $R$ such that the post flare flux at $t=9643M$ is fixed to 0.2 Jy 
    (this is the 230 GHz flux for the $R=1$ model at this time when the density normalization is chosen so that the time-averaged 230 GHz flux equals to 0.66 Jy),
    to highlight how the flux evolution during the flare depends on electron temperature model.
    \textit{Dimming} at 230 GHz occurs with $R \equiv T_p/T_e \lesssim 20$ while we see a \textit{brightening} and then fading for $R \gtrsim 20$.}
    \label{fig:flux}
\end{figure*}

\section{Light Curves at 230 GHz}

\label{sec:lightcurve}

In this section, we study the observational signatures of the flare state in 230 GHz light curves.
As argued in \citet{ripperda2022black}, the dissipation of the jet's magnetic energy through transient reconnection events near the event horizon is a possible mechanism to power observed flares from black holes.
We find three major energetic reconnection events between $t\!=\!5,\!000M$ and $t\!=\!10,\!000M$, indicated by the decay of the magnetic flux $\tilde{\phi}_{\rm horizon} \equiv \frac{1}{2} \int_0^{2\pi} \int_0^{\pi} \left| ^*F^{rt}\right|\sqrt{-g}{\rm d}\theta{\rm d}\phi$ on the horizon, which are highlighted by the grey bands in the left panel of Figure \ref{fig:fluid}.
In the right panel, we confirm that the maximum fluid temperature (defined in Equation \ref{eq:R}) does increase when magnetic reconnection happens, due to the electromagnetic energy converted to heat by the reconnection.
Note that the plasma was modelled as a single-temperature thermal fluid in the GRMHD simulation, whereas it is very likely that the electrons around realistic black holes are non-thermal and have a different energy distribution than the protons \citep[e.g.][]{chael2017evolving}.
Therefore, the maximum electron 
energy is likely larger than that associated with the maximum temperature shown in Figure \ref{fig:fluid}.
Indeed, according to \citet{Sironi2014Relativistic,hakobyan2022radiative}, particle acceleration in reconnection at high $\sigma$ is particularly efficient in that a large fraction of the dissipated energy ends up in high energy particles. In addition, the high energy particles cool rapidly by synchrotron radiation so high energy flares appear likely if reconnection is sourced by highly magnetized plasma,
as suggested by GRMHD simulations. Thus we are motivated to refer to magnetic flux eruptions as flares throughout this paper.  Here we will use $T_{\max}$ as a proxy for the high energy emission, since direct, self-consistent modelling of the X-ray or $\gamma$-ray light curves is beyond the scope of this work. {We note that $T_{\rm max} \sim \sigma_{\rm floor}$ depends directly on the magnetization in the jet, which is set by $\sigma_{\rm floor}=25$ in our GRMHD simulation. We will discuss the implications of $\sigma_{\rm floor}$ being much smaller than realistic values in Section \ref{sec:discuss}.}

\begin{table*}
	\centering
	\def\arraystretch{1.1}
	\caption{Where does the majority of the emission come from?
	For each fluid quantity $x$, we find $x_{\rm low}$ such that if we ignore the region with $x>x_{\rm low}$, the total 230 GHz flux drops to 15\% of the total value, and similarly for $x_{\rm high}$.
	The region with $x\in(x_{\rm low}, x_{\rm high})$ thus contributes 70\% of the flux, in the limit of negligible absorption.
	For the latitude $\tilde{\theta}$ we only reports $|\tilde{\theta}|_{\rm high}$ since most of the emission comes from the equatorial disk.
    Note that $n_e$ and $B$ depends on the density normalization factor between physical and simulation units, which is $4.88\,(156)$ times larger for $R=10\,(100)$ compared with $R=1$.}
	\label{tab:major}
	\begin{tabular}{cccccccc} 
		\hline
		$t$ / $M$ & ($R_{\rm low}$, $R_{\rm high}$) & ($n_{\rm e,low}$, $n_{\rm e,high}$) / cm$^{-3}$ & ($T_{\rm e,low}$, $T_{\rm e,high}$) / K & ($\beta_{\rm low}$, $\beta_{\rm high}$) & ($B_{\rm low}$, $B_{\rm high}$) / G & ($r_{\rm min}$, $r_{\rm max}$) / $r_g$ & $|\tilde{\theta}|_{\rm high}$ / $^{\circ}$ \\
		\hline
		\multirow{3}{*}{9113} & (1, 1) & ($2.96\times10^3$, $2.53\times10^4$) & ($1.94\times10^{11}$, $5.87\times10^{11}$) & (0.29, 3.47) & (2.26, 8.65) & (2.99, 6.58) & 15.3 \\
		& (10, 10) & ($2.29\times10^4$, $2.07\times10^5$) & ($6.29\times10^{10}$, $2.24\times10^{11}$) & (0.32, 4.65) & (8.66, 31.1) & (2.26, 4.60) & 14.3  \\
		& (100, 100) & ($4.13\times10^5$, $9.25\times10^6$) & ($9.25\times10^{9}$, $9.49\times10^{10}$) & (0.30, 14.8) & (50.2, 258) & (1.81, 4.34) & 15.8 \\
		\hline
		\multirow{3}{*}{9378} & (1, 1) & ($6.50\times10^2$, $1.53\times10^4$) & ($2.31\times10^{11}$, $7.32\times10^{11}$) & (0.44, 3.88) & (1.21, 5.55) & (3.11, 9.06) & 23.9 \\
		& (10, 10) & ($1.49\times10^3$, $8.38\times10^4$) & ($8.18\times10^{10}$, $3.76\times10^{11}$) & (0.54, 4.39) & (2.80, 17.8) & (2.59, 7.79) & 22.6  \\
		& (100, 100) & ($4.08\times10^3$, $3.05\times10^5$) & ($3.38\times10^{10}$, $1.92\times10^{11}$) & (1.11, 10.1) & (6.29, 51.3) & (3.32, 15.2) & 33.2 \\
		\hline
		\multirow{3}{*}{9643} & (1, 1) & ($9.66\times10^2$, $9.00\times10^3$) & ($1.85\times10^{11}$, $5.74\times10^{11}$) & (0.32, 3.50) & (1.24, 4.95) & (2.93, 7.63) & 23.7 \\
		& (10, 10) & ($9.56\times10^3$, $6.13\times10^4$) & ($7.17\times10^{10}$, $2.02\times10^{11}$) & (0.41, 3.40) & (5.32, 17.4) & (2.23, 4.36) & 14.2  \\
		& (100, 100) & ($2.25\times10^5$, $2.10\times10^6$) & ($1.48\times10^{10}$, $6.40\times10^{10}$) & (0.53, 5.01) & (35.8, 125) & (1.88, 3.78) & 14.3 \\
		\hline
	\end{tabular}
\end{table*}

\begin{figure*}
    \centering
	\includegraphics[width=15cm]{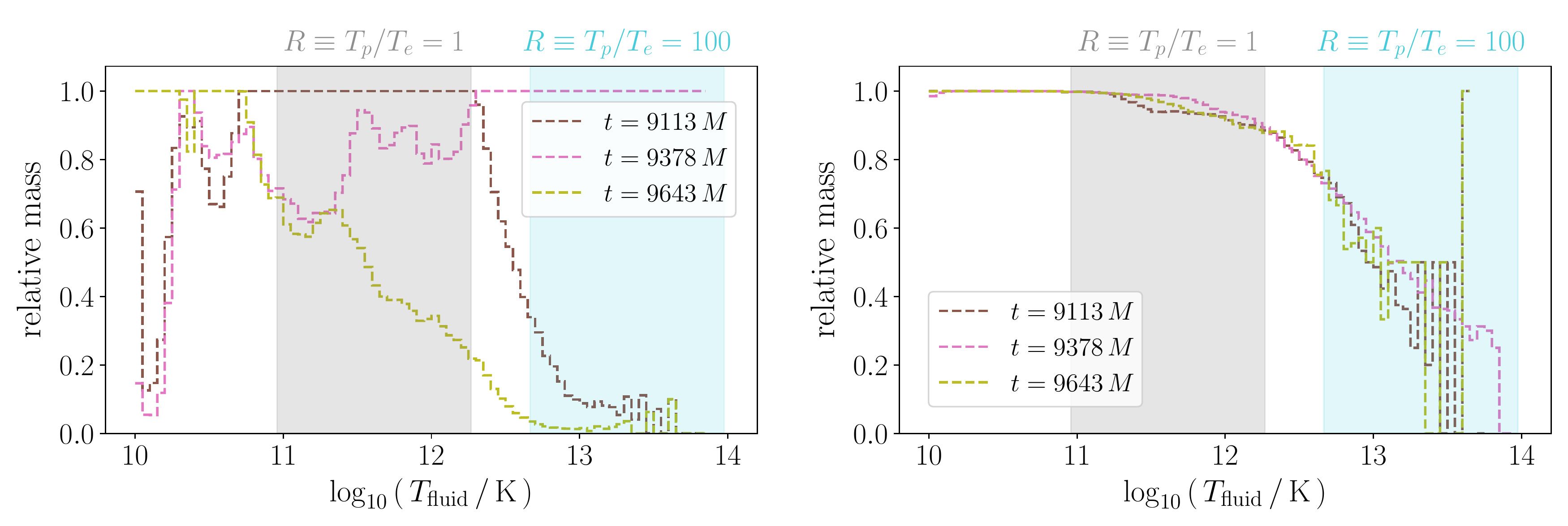}
    \caption{The $T_{\rm fluid}$ distribution of the plasma in the pre-flare $t=9113M$, mid-flare $t=9378M$ and post-flare $t=9643M$ snapshots (see the left panel of Figure \ref{fig:fluid}).
    We only include the region with $r<20$, latitude $\left|\tilde{\theta}\right|<60^{\circ}$, which is responsible for the majority of 230 GHz emission according to Table \ref{tab:major}.
    {\em Left:} using $\sigma_{\rm cut}=1$, we show the relative mass of plasma in each bin normalized by the maximum value among the three snapshots.
    The mid-flare $t=9378M$ state has the highest temperature $T_{\rm fluid} \gtrsim 2\times10^{12}\,$K plasma, while the pre-flare $t=9113M$ state has the most intermediate temperature $7\times10^{10}\,$K$\,\lesssim T_{\rm fluid}\lesssim2\times 10^{12}\,$K plasma.   The synchrotron emission at 230 GHz is dominated by plasma with $T_e \sim 10^{11}$ K.   
    The red and blue bands show the characteristic GRMHD fluid temperature associated with this synchrotron-emitting plasma for $R\equiv T_p/T_e = 1$ and 100, respectively.  High electron temperature models (e.g., R = 1) produce 230 GHz dimming during flux eruptions (Fig. \ref{fig:flux}) because the plasma is too hot to radiate effectively at 230 GHz; by contrast, low electron temperature models (R = 100) brighten during flux eruptions because the high $T_{\rm fluid}\gtrsim 2\times10^{12}\,$K plasma has lower $T_e$ compared with other electron temperature models and thus radiates efficiently at 230 GHz.  {\rm Right:} we show the ratio of the total plasma mass with $\sigma_{\rm cut}=1$ to the total plasma mass with $\sigma_{\rm cut}=10$.
    Higher temperature $T_{\rm fluid}\gtrsim 5\times10^{12}\,$K regions are more likely to have larger $\sigma\sim\sigma_{\rm cut}$.
    Therefore, ray tracing images with higher $R\gtrsim100$ are more sensitive to the choice of $\sigma_{\rm cut}$ (Section \ref{sec:sigmacut}).}
    \label{fig:TL}
\end{figure*}

Figure \ref{fig:nT} visualizes equatorial density and temperature fields for the $t=9113M$ pre-flare quiescent, $t=9378M$ mid-flare, and $t=9643M$ post-flare quiescent states (labeled in Figure \ref{fig:fluid}).
In the $t=9378\,M$ mid-flare state we see reconnection-heated hot $\gtrsim 5\times10^{12}\,$K plasma out to $\sim15 r_g$.
How does the 230 GHz emission change during the high energy flares?
In the left panel of Figure \ref{fig:flux}, we plot the 230 GHz light curves for six electron temperature models that are similar to those used in EHT analysis \citep{eht2019m87v}.
We find two different patterns of 230 GHz light curves, depending on the $T_e$ model. 
For all but the lowest electron temperature model $R=R_{\rm high}=R_{\rm low}=100$, there is a strong correlation between $\tilde{\phi}_{\rm horizon}$ and 230 GHz emission: as $\tilde{\phi}_{\rm horizon}$ drops during the flares, the 230 GHz flux also reduces by up to 80\%, in contrast to the expected brightening at higher energy bands.
With $R=100$, however, the synchrotron flux at 230 GHz increases simultaneously with $-\,{\rm d}\tilde{\phi}_{\rm \,horizon}\,/\,{\rm d}t$ (and therefore $T_{\max}$), meaning that the high energy flare would be accompanied by a 230 GHz counterpart.

In the right panel of Figure \ref{fig:flux}, we calculate the 230 GHz flux with different $R=R_{\rm low}=R_{\rm high}$ between 1 and 500 for the three times identified in Figure \ref{fig:fluid} and shown in Figure \ref{fig:nT}.   
Since here we want to compare the relative strength of emission at the three times, we adjust the density normalization such that the 230 GHz flux at $t=9643M$ is fixed to 0.2$\,$Jy for all the models.  This facilitates easy comparison of the pre and mid-flare emission relative to the post-flare emission (note that this normalization choice is such that the time-averaged 230 GHz flux for the $R=1$ model is the fiducial 0.66 Jy).    We find similar results as the left panel of Figure \ref{fig:flux}: with smaller $R\lesssim 20$, the 230 GHz flux drops monotonically as $\tilde{\phi}_{\rm horizon}$ decays.
On the other hand, when $R\gtrsim 20$, the 230 GHz flux of $t=9378M$ mid-flare state exceeds the $t=9113M$ pre-flare state: the flare \textit{dimming} at 230 GHz with small $R$ models eventually turns into flare \textit{brightening} with large $R$ models, in accordance with the lightcurve predictions in the left panel of Figure \ref{fig:flux}.

The numerical results in Figure \ref{fig:flux} can also be understood analytically using the well-understood properties of synchrotron emission; we assume optically thin emission in what follows.    Plasma with dimensionless temperature $\theta_e = kT_e/m_e c^2 = 10 \theta_{10}$, i.e., $T_e \simeq 6 \times 10^{10} \theta_{10}$ K in a magnetic field of strength $B = 10 B_{10}$ G emits synchrotron most efficiently, i.e., the emissivity $\nu j_\nu$ peaks, at a frequency 
\begin{equation}
\nu_{\rm peak} \sim 5 \frac{e B}{m_e c} \theta_e^2 \simeq 90 \, B_{10} \, \theta_{10}^2 \ {\rm GHz}.
\end{equation}
For $\nu \ll \nu_{\rm peak}$ the synchrotron emission scales with plasma parameters as $j_\nu \propto n B^{3/4} \theta_e^{-1/2}$.   This shows that for high electron temperature models in which mm observations are at $\nu_{\rm obs} = 230$ GHz $\lesssim \nu_{\rm peak}$, the emission at 230 GHz will decrease with the increasing electron temperature during a flux eruption, as seen in the lower $R$ models in Figure \ref{fig:flux}.
For $\nu \gg \nu_{\rm peak}$, on the other hand, as is the case in models with lower electron temperatures, the synchrotron emission scales with plasma parameters as $j_\nu \propto n \theta_e^{-2} \exp[-5.5(\nu/\nu_{\rm peak})^{1/3}]$.   The exponential dependence on $T_e^{-2/3}$ implies that when the electron temperature is low the emission at 230 GHz increases with increasing electron temperature.   The mm synchrotron emission will thus increase during a flux eruption, as seen in the higher $R$ models in Figure \ref{fig:flux}.    More generally, synchrotron emission at 230 GHz is particularly sensitive to electrons with temperatures {corresponding to emission at} $\nu_{\rm peak} \sim 230$ GHz, i.e., $\theta_e \simeq 16 B_{10}^{-1/2}$.  During a flux eruption whether the mm synchrotron flux increases or decreases thus depends on the details of electron heating for plasma with $T_e \simeq 10^{11}$ K (which is much less than the characteristic fluid temperatures reached during the eruption in GRMHD simulations; see Figure \ref{fig:nT}).

\begin{figure*}
	\includegraphics[width=\textwidth]{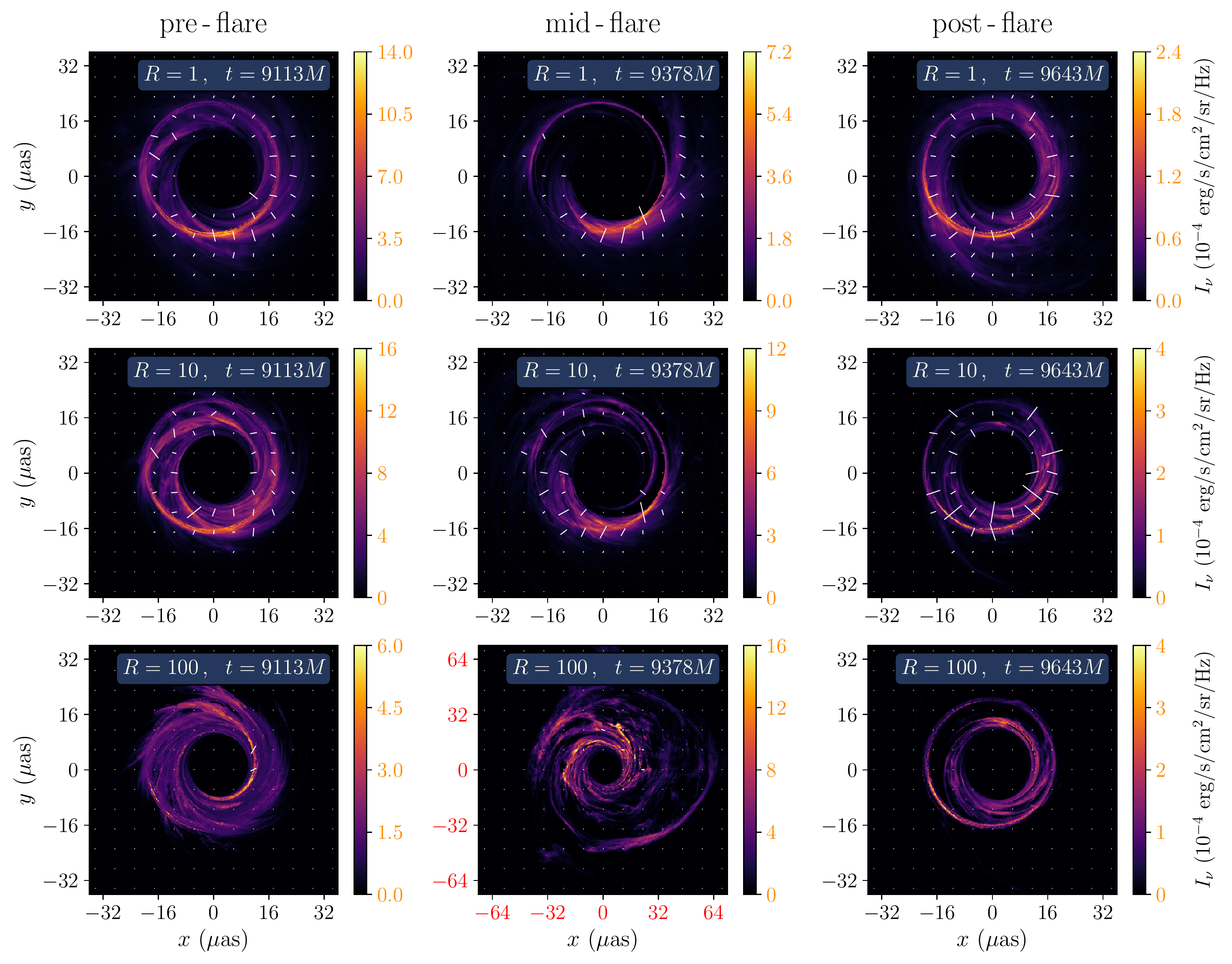}
    \caption{The intensity and polarization maps for three $R=R_{\rm low}=R_{\rm high}$ models at three snapshots.
    The tick direction represents the direction of linear polarization, while the tick length is proportional to $\sqrt{\mathcal{Q}^2+\mathcal{U}^2}$.
    {\color{red} Note the red} {\color{orange} and orange} {\color{red} labels which are different between different panels.}
    For $R=1$ or 10, the 230 GHz flux drops during the flares but the ring morphology does not change significantly. For $R=100$, however, the emission region becomes much more spatially exended at $t=9378M$ during the flare.}
    \label{fig:polar-07-00}
\end{figure*}

\begin{figure*}
	\includegraphics[width=\textwidth]{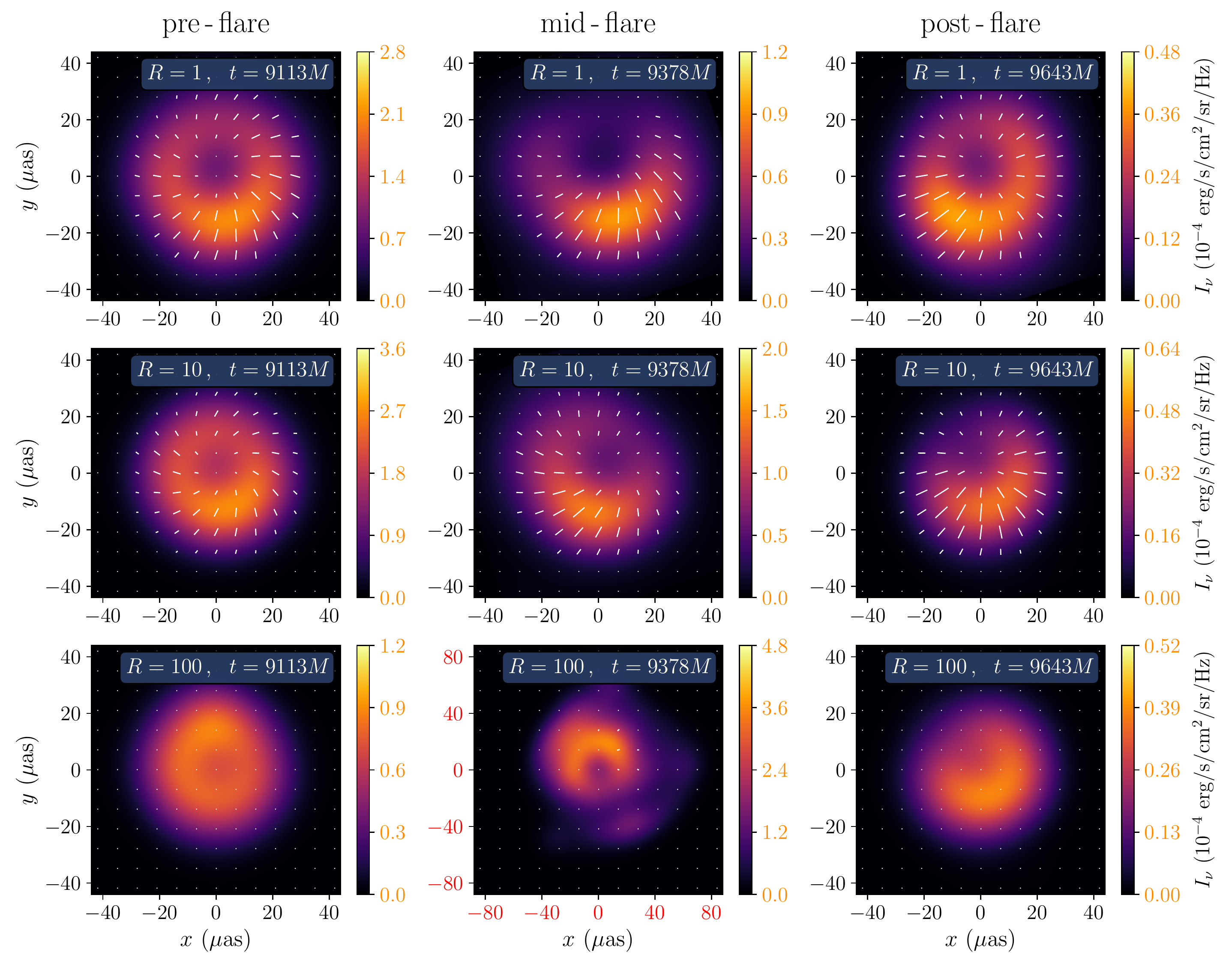}
    \caption{Similar to Figure \ref{fig:polar-07-00}, but blurred with a 20$\,\mu$as FWHM Gaussian kernel, which matches the current EHT resolution.
    The polarization pattern does not change much for $R=1$ or 10, but for $R=100$ it becomes noticeably more polarized at $t=9378M$.}
    \label{fig:polar-07-01}
\end{figure*}

\begin{figure*}
	\includegraphics[width=\textwidth]{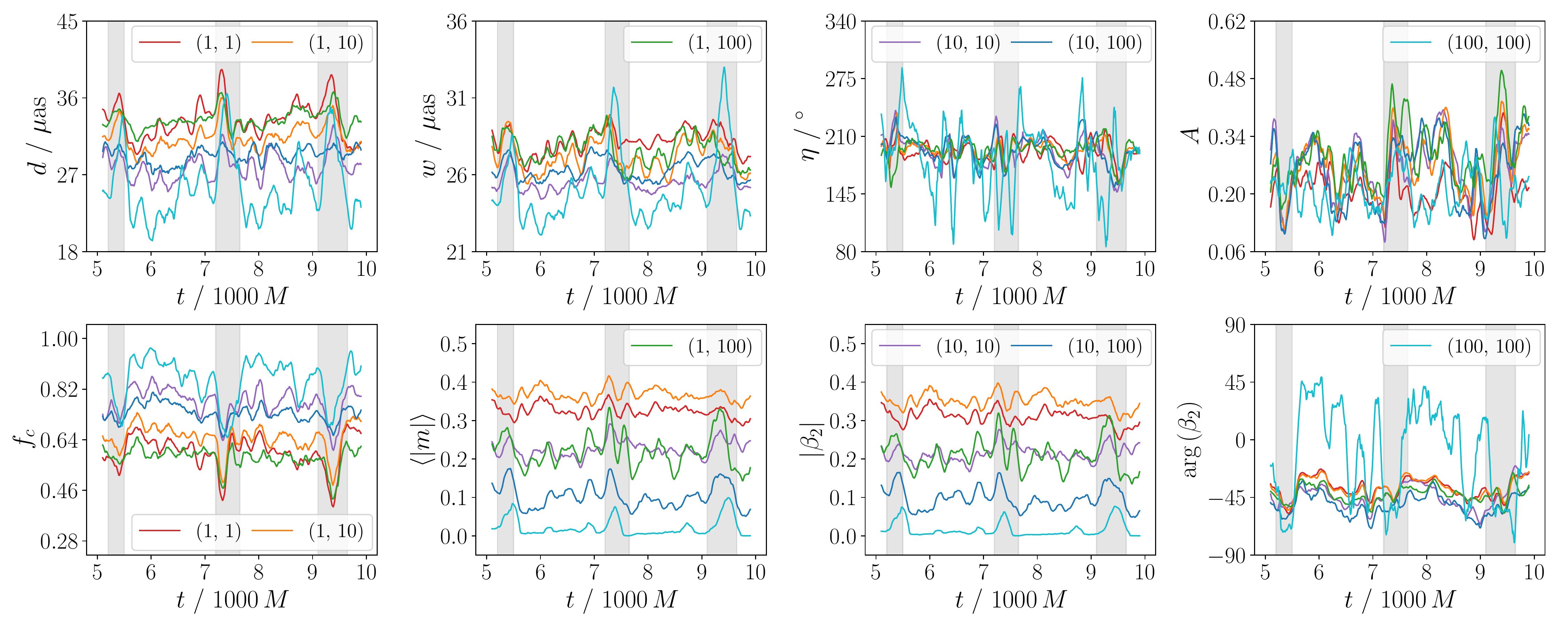}
    \caption{Image statistics as a function of time for five different ($R_{\rm low}, R_{\rm high}$) electron temperature models, smoothed by a moving average with a window of 150$\,M$; see Section \ref{sec:method} for the exact definition of the statistics.
    The grey bands indicate the three major flare states.
    During the flares, we see an increase of ring diameter $d$, width $w$ and degree of asymmetry $A$, and a decrease of fractional central brightness $f_C$, due to the ejection of the inner disk.
    The other statistics are similar between flare and quiescent states.}
    \label{fig:stats}
\end{figure*}

\begin{figure*}
	\includegraphics[width=15cm]{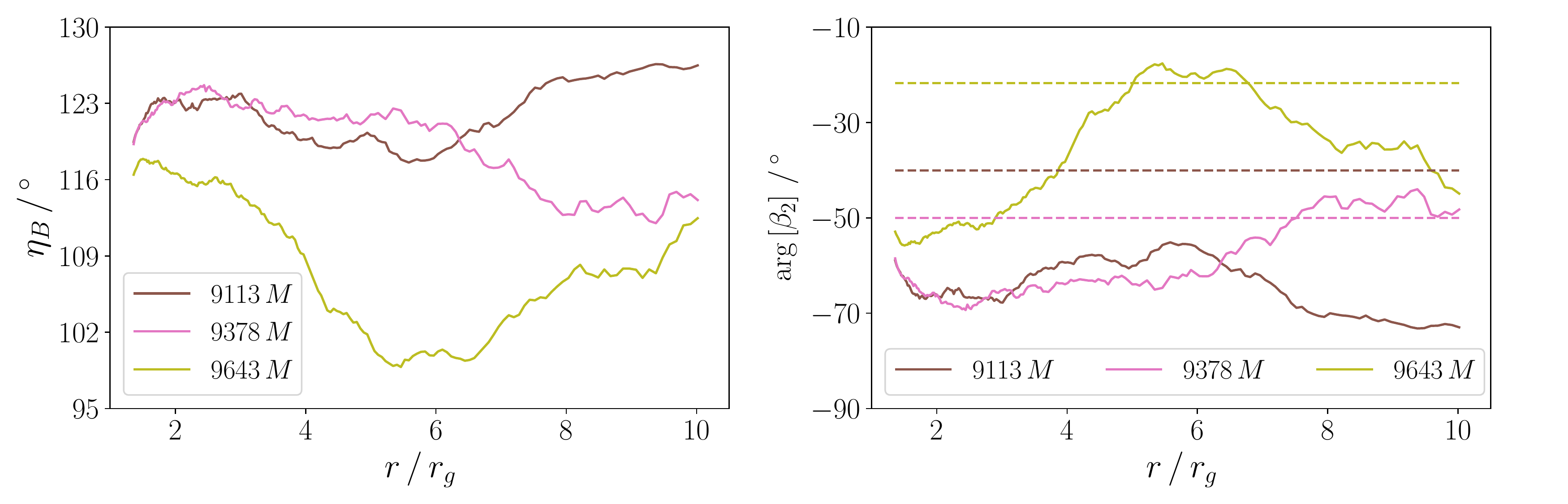}
    \caption{{\em Left:} the mean direction of the equatorial magnetic field $\eta_B$, as defined in Equation \ref{eq:etaB}.
    $\eta_B$ does not change significantly with time, neither is it sensitive to the radius within $\sim10\,r_g$.
    {\em Right:} dashed lines represent the actual image-averaged arg$[\beta_2]$ from the $R_{\rm low}=R_{\rm high}=1$ images, which is close to the results of all low R models (see Figure \ref{fig:stats}).
    Solid lines show the semi-analytic computation of arg$[\beta_2]$ in Equation \ref{eq:argbeta2}, which is a reasonable approximation of the actual measured arg$[\beta_2]$ from ray tracing images.}
    \label{fig:beta2}
\end{figure*}


To better understand the dependence of the 230 GHz light curves on the electron temperature models, we ray-trace with various cuts (i.e. ignore certain regions of plasma based on different fluid quantities) to identify the characteristic fluid quantities in the emission region; we find the lower and upper bounds for the fluid quantities that are responsible for the majority of the emission, and list the results in Table \ref{tab:major}.
While the characteristic electron temperature $T_e$ becomes lower {for} larger $R$ models, the characteristic fluid temperature $T_{\rm fluid}$ actually increases with $R$ (see Equation \ref{eq:R}).
Roughly speaking, for $R=1$, 10 and 100, the characteristic $T_{\rm fluid}$ for 230 GHz emission is approximately $4\times10^{11}\,$K, $8\times10^{11}\,$K and $4\times10^{12}\,$K, respectively, which does not change significantly between the snapshots.
Therefore, for different $R$ models the 230 GHz emission comes from different parts of the accretion flow (which we will specify further next) with different $T_{\rm fluid}$, and thus may have different time evolution.

In the left panel of Figure \ref{fig:TL}, we plot the relative plasma mass within different $\log_{10} T_{\rm fluid}$ bins for the three snapshots.
The $t=9378M$ mid-flare state has the most high temperature $T_{\rm fluid} \gtrsim 2\times10^{12}\,$K plasma, due to the energy released by reconnection in the equatorial current sheet, which is consistent with the right panel of Figure \ref{fig:fluid} where we use $T_{\max}$ as a proxy of the mass of high temperature plasma.  
On the other hand, the mass of intermediate temperature $7\times10^{10}\,$K$\,\lesssim T_{\rm fluid}\lesssim2\times 10^{12}\,$K plasma keeps decreasing during the flare state, which we attribute to the evacuation of the inner accretion disk \citep[see Figure 1 of][]{ripperda2022black}.
We also show the characteristic $T_{\rm fluid}$ for 230 GHz emission for $R=1$ and $100$ with the red and blue bands, defined as the range of $T_{\rm fluid}$ in which the 230 GHz synchrotron emissivity is larger than half of the peak value over all $T_{\rm fluid}$, assuming $B=20\,$G and fixed $n_e$.
The locations of red and blue bands  move somewhat if we choose e.g. $B=5\,$G or $50\,$G; this does not, however, change our main conclusions.   

The left panel of Figure \ref{fig:TL}  elucidates the strong connection between the electron temperature model and the correlation or anti-correlation of the mm synchrotron flux with the magnetic flux eruption.   For electron temperature models with $R \sim 1-10$ the high temperature plasma created during the flux eruption does not radiate effectively at 230 GHz which is why the flux eruption is accompanied by a decreasing 230 GHz flux.     By contrast, for electron temperature models with $R \sim 30-100$, the high $T_{\rm fluid}$ plasma has just the right electron temperature to emit significantly at 230 GHz.  This is why the flux eruption is accompanied by an increased 230 GHz flux in higher $R$ electron models.

The right panel of Figure \ref{fig:TL} shows the ratio of the total plasma mass with $\sigma_{\rm cut}=1$ to the total plasma mass with $\sigma_{\rm cut}=10$.  The regions with trans-relativistic $\sigma\sim\sigma_{\rm cut}$ usually also have higher $T_{\rm fluid}$, so one would expect that ray tracing calculations at higher frequencies or with lower electron temperature models (with large $R$ in Equation \ref{eq:R}), for which the emission is from the regions with higher $T_{\rm fluid}$, are more sensitive to the choice of $\sigma_{\rm cut}$.   This is consistent with our convergence calculations in Appendix \ref{sec:sigmacut}.

\section{Images at 230 GHz}

\label{sec:image}

We show the intensity and polarization maps for three $T_e$ models and the three typical snapshots in Figures \ref{fig:polar-07-00} and \ref{fig:polar-07-01}.
\footnote{We note that the polarization ticks in Figures 5, 7 and 11 of \citet{jia2022observational} are not correctly plotted, although their quantitative results for the $m$ and $\beta_2$ statistics are not affected by this issue.}
For simplicity, we present three models with $R_{\rm low}=R_{\rm high}$, while we find that the images with $R_{\rm low}<R_{\rm high}$ are generally similar to images with $R_{\rm low}=R_{\rm high}=R^*$ where the effective $R^*$ lies between $R_{\rm low}$ and $R_{\rm high}$.   Note that the $R = 100$ model mid-flare ($t = 9378 M$) image is approximately 4 times larger than the other panels in terms of area.

As with the light curves, we find two different regimes for 230 GHz images depending on the electron temperature model.   For higher $T_e$ models with $R=1$ or 10, we only see a steady decline of the 230 GHz flux, but the ring morphology does not change much during the flares.
As $R$ increases, the quiescent state emission tends to move inwards, as the 230 GHz emission for larger $R$ models comes from the region with lower $T_e$ but higher $T_{\rm fluid}$.
With $R=100$, the flare image changes significantly compared with quiescent states, since the hot electrons produced at larger radii $r\gtrsim 5r_g$ during the flares dominate the 230 GHz emission.
We note that not only the characteristic radii of the emission increases, but also the emission extends to larger $\left|\tilde{\theta}\right|$ (Table \ref{tab:major}) and has contributions from both the current sheet and the heated jet sheath from reconnection exhaust \citep{ripperda2022black}, implying that thin disk semi-analytic models may no longer be suitable for modelling the emission for such cases.
Comparing with the density and temperature maps in Figure \ref{fig:nT}, the $T_{\rm fluid}\gtrsim 5\times10^{12}\,$K hot flow at $t=9378\,M$ is only visible in the image with $R=100$, since with lower $R\lesssim10$ it will be too hot to contribute significantly to the 230 GHz emission.

In Figure \ref{fig:stats}, we compute the 230 GHz image statistics introduced in Section \ref{sec:method}.
As we already concluded from the total intensity images, the colder electron models (larger $R$) generally show a smaller and thinner ring, since the emission comes from the inner regions with higher $T_{\rm fluid}$ (but lower $T_e$).
During the flares, the ring diameter $d$ and width $w$ increase while the fractional central brightness $f_C$ decreases, as the ejection of the inner disk moves the luminous plasma farther from the black hole.
For higher electron temperature models with $R\lesssim10$, the ring orientation $\eta$ does not change much since the ring asymmetry is mainly due to Doppler beaming of the accretion flow.
On the other hand, for $R=100$, the emission region is more extended and the ring asymmetry mainly comes from the asymmetric distribution of hot plasma spiraling down into the black hole, which leads to a larger variation of the ring orientation.
Larger $R$ models generally produce less polarized images, as they need a larger fluid density to match the observed 230 GHz flux which enhances Faraday depolarization.
As the fluid density drops in the region ($\sim$ inner $10 r_g$) where the accretion disk is ejected, during the flare states, the 230 GHz emission also becomes more polarized.

\begin{figure*}
	\includegraphics[width=15cm]{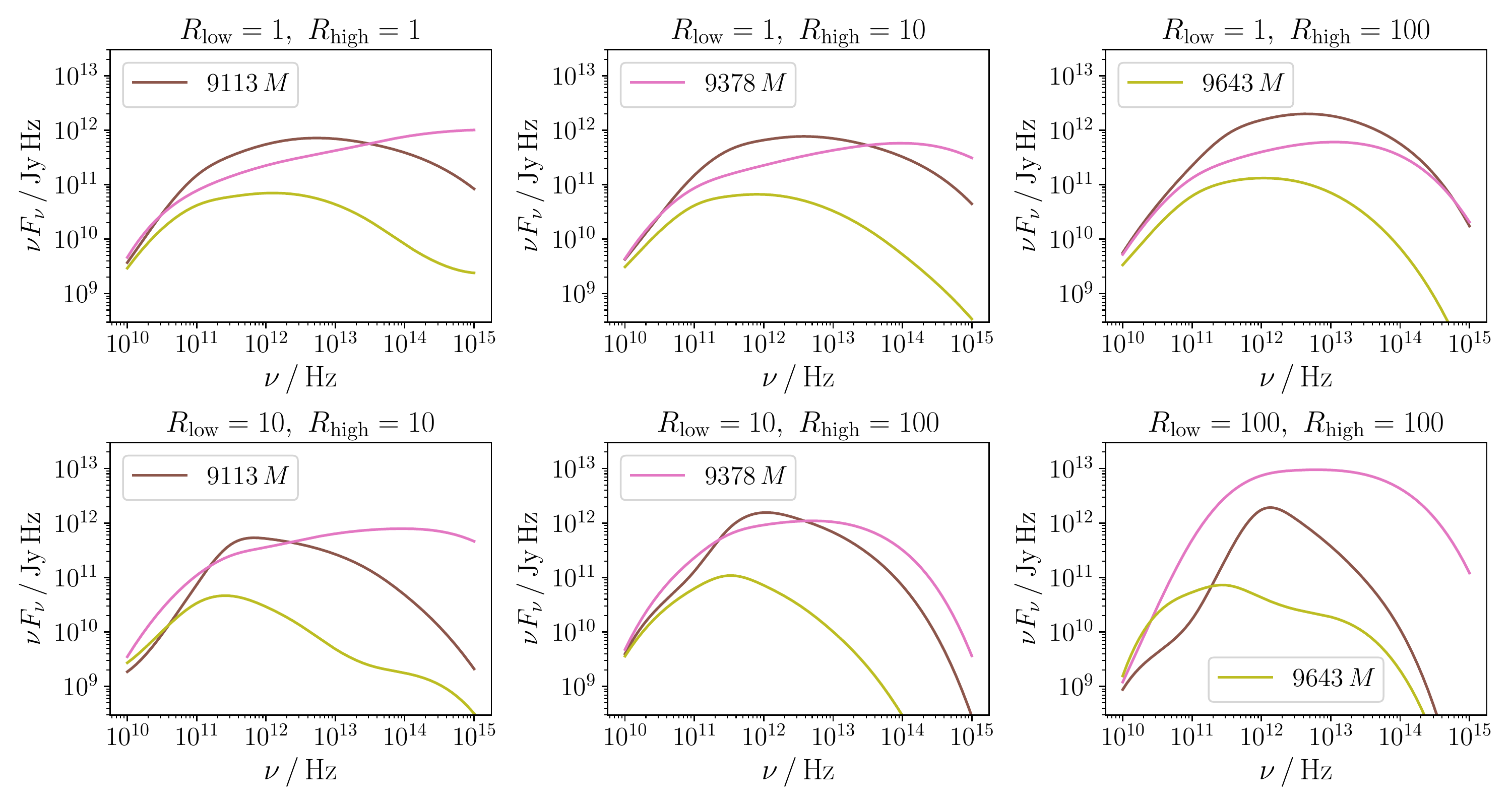}
    \caption{Sychrotron emission spectra from radio to UV for the $t=9113M$ pre-flare, $t=9378M$ mid-flare and $t=9643M$ post-flare states.
    The mid-flare state is generally the brightest at higher frequencies, however we note that non-thermal electrons and inverse Compton emission should be properly modeled for a quantitative prediction of the higher frequency spectra.}
    \label{fig:spec}
\end{figure*}

While magnetic reconnection changes the topology of the magnetic field during the flare states \citep{ripperda2022black}, the mean direction of the equatorial magnetic field,
\begin{equation}
    \eta_B(r) \equiv \left< \arctan(\dfrac{\sqrt{g_{\phi\phi}}\,|B^{\phi}|}{\sqrt{g_{rr}}\,B^{r}{\rm sign}(B^{\phi})}) \right>_{\theta\in(75^{\circ},\,105^{\circ}),\,\phi\in(0^{\circ},\,360^{\circ})},
    \label{eq:etaB}
\end{equation}
does not change substantially with time, as shown in the left panel of Figure \ref{fig:beta2}.
Here the range of the arctan function is set to $[0^{\circ},\,180^{\circ})$, and $\eta_B$ is invariant under a sign inversion of the magnetic field since the synchrotron emissivity is also unchanged.
According to Equation 39 in \citet{narayan2021polarized}, the leading order prediction of arg$[\beta_2]$ for optically thin, axis-symmetric, equatorial plasma and magnetic field profile is given by
\begin{equation}
    {\rm arg}[\beta_2] \simeq \pi - 2\,\eta_B,
    \label{eq:argbeta2}
\end{equation}
for face-on observers from the south pole direction.
This indeed gives a reasonable approximation of the actual arg$[\beta_2]$, as shown in the right panel of Figure \ref{fig:beta2} (the agreement is somewhat worse at the pre-flare time, which we don't have a simple explanation for).
Therefore, for all the low $R$ electron temperature models which lead to optically thin 230 GHz synchrotron emission, arg$[\beta_2]$ does not change significantly with time, neither is it sensitive to the exact values of $R_{\rm low}$ and $R_{\rm high}$.
On the other hand, the 230 GHz synchrotron emission becomes Faraday thick for large $R$ models, for which Equation \ref{eq:argbeta2} no longer holds.
In this case, we find strong Faraday depolarization (the intensity-weighted Faraday rotation depth $\left<\tau_{\rho_V}\right>\sim5000$ for $R_{\rm low}=R_{\rm high}=100$) during the quiescent state, such that $|\beta_2|$ is small and arg$[\beta_2]$ deviates from the predictions of Equation \ref{eq:argbeta2}.
During the flare state, the plasma density drops and so does the Faraday rotation depth.
Therefore, $|\beta_2|$ increases and arg$[\beta_2]$ becomes closer to the optically-thin limit in Equation \ref{eq:argbeta2}.

We note that here $\arg[\beta_2]$ is inconsistent with EHT measurements $197^{\circ}\leq\arg(\beta_2)\leq231^{\circ}$ \citep{eht2021m87viii} for all the snapshots and electron temperature models, implying that the magnetic field is probably too azimuthal at $r\lesssim5r_g$ in the large spin MAD simulations \citep{narayan2021polarized}.   This is consistent with previous theoretical work, e.g., Figure 28 of \citet{eht2021m87viii}).

\section{Multi-wavelength Synchrotron Spectra}

\label{sec:multi}

In this section, we go beyond 230 GHz and compute the synchrotron emission spectra from $10^{10}$ to $10^{15}$ Hz; the results are shown in Figure \ref{fig:spec}, with each curve interpolated between 11 different frequencies.
We use the same density normalization as in the previous sections, namely the time averaged 230 GHz flux should be 0.66 Jy to match EHT observations.
Generally, the $t=9378M$ mid-flare state has the largest flux at higher frequencies ($\gtrsim\!10^{13}$ Hz), since there is more plasma heated up by reconnection in the current sheets.   Note that this implies IR and even higher-energy "flares" associated with flux eruptions nearly independent of whether the mm brightens or fades (the exception is the $R_{\rm low}=1, R_{\rm high} = 100$ model). 
The spectra in Figure \ref{fig:spec} drop faster at higher frequencies for lower electron temperature models, as there are not many electrons that are hot enough to emit at such frequencies.

In this calculation we only include thermal electrons and synchrotron emission, whereas a better modelling of non-thermal electrons, pair production and thermal and non-thermal inverse Compton {scattering} is required for a quantitative analysis of the spectra at higher frequencies {(X-ray and $\gamma$-ray, and even optical-IR for M87*)}; such modeling is is beyond the scope of this paper (see, e.g., \citealt{Ryan2018,hakobyan2022radiative} for work including more of the relevant radiative processes).
Another uncertainty comes from the choice of $\sigma_{\rm cut}$: unlike the 230 GHz computations, we find that the high frequency ray tracing results are sensitive to $\sigma_{\rm cut}$, since the $\sigma\sim\sigma_{\rm cut}$ region may have higher temperature and thus dominate the emission at $\gtrsim\!10^{13}$ Hz.
Here we use $\sigma_{\rm cut}=10$ for the spectra, in contrast to $\sigma_{\rm cut}=1$ for 230 GHz emission in the previous sections.
We find that with $\sigma_{\rm cut}=1$ and lower $T_e$ models, the emission at $\gtrsim\!10^{14}$ Hz basically drops to zero for many snapshots, as all the electrons that are hot enough to emit at such high frequencies are removed by $\sigma_{\rm cut}=1$.
Nevertheless, Figure \ref{fig:spec} does confirm the qualitative trend that the high frequency flux increases during the flare state, which is the origin of the observed bright \textit{flares}.

\section{Discussion}

A promising source of high energy flares in accreting black holes such as Sgr A*, M87*, and related systems is  reconnection in near-horizon current sheets.  Such reconnection is particularly prominent and energetically important during magnetic flux eruptions in MAD accretion models.   This model is attractive because it qualitatively explains the timescales and duty cycles of the observed flares as well as many aspects of the observed radiation \citep[e.g.][]{Dodds-Eden2010Time-Dependent,Dexter2020Sgr,Porth2021Flares}.  Intriguingly, this model also associates the flares with a dynamically critical aspect of the theoretical model, namely the {magnetic flux eruptions and the associated magnetic energy} dissipation 
required for accretion to continue in spite of the energetically dominant magnetic energy in the system.    The main goal of this paper has been to study the 230 GHz emission associated with the same magnetic flux eruptions posited to produce the high energy flares.

Should the high energy flares in fact be accompanied by a mm counterpart?    The short answer is \textit{maybe}, depending on how efficiently electrons with Lorentz factors of $\sim 100$ (temperatures of $\sim 10^{11}$ K) are heated during the flare.   In the context of GRMHD modeling like that employed in this work, this depends on the relation between $T_e$ and $T_{\rm fluid}$ (the GRMHD simulation temperature) which unfortunately is still poorly understood.  The energy released by reconnection heats up the plasma to $T_{\rm fluid}\gtrsim 2\times10^{12}\,$K.
However, since the accretion flow in low-luminosity AGN very likely has different electron temperatures than proton temperatures, one needs a prescription for the electron temperature that determines the synchrotron emissivity.
With higher $T_e$ models like $R\equiv T_p/T_e=R_{\rm low}=R_{\rm high}=1$, we find that the $T_{\rm fluid}\gtrsim 2\times10^{12}\,$K hot plasma contributes negligible emission at 230 GHz, and the 230 GHz emission gradually \textit{dims} during the flare, although at higher frequencies the synchrotron flux does increase.
We have also studied the spatially resolved emission during the flares in the context of future EHT observations.   We find that for models in which the 230 GHz emission dims, the  diameter of the high surface brightness "ring" of emission slightly increases and the fractional central brightness decreases, due to the ejection of the inner accretion disk.    The polarization is similar between quiescent and flare states because the equatorial magnetic field direction does not change much during the flares (Figure \ref{fig:beta2}).

On the other hand, with lower $T_e$ models like $R\equiv T_p/T_e=R_{\rm low}=R_{\rm high}=100$, the flare-state hot electrons are just the right temperature to emit 230 GHz synchrotron radiation, which leads to a 230 GHz flare \textit{brightening}, simultaneous with the high energy flares.
Such hot electrons typically have a broader spatial distribution out to $r\lesssim15 r_g$ (Figure \ref{fig:nT}), so that the size of the ring also increases significantly.
The polarization fraction increases while the orientation of the polarization (arg$[\beta_2]$) fluctuates during the flares, which we attribute to the evacuation of the inner accretion disk leading to less Faraday depolarization.  It is important to stress that the mm brightening for $R = 100$ models found here may depend on the magnetization ceiling of $\sigma_{\rm floor} = 25$ used in the GRMHD simulation.   Higher $\sigma_{\rm floor}$ implies a higher temperature $T_{\rm fluid}$ of reconnection-heated plasma (and vice-versa; see \citealt{Ripperda2020}) and so the appropriate value of $R$ that corresponds to the transition between mm brightening and dimming during flares likely increases with increasing  $\sigma_{\rm floor}$ (such that $T_e \sim 10^{11}$ K).
We also note that models with $R=R_{\rm low}\sim R_{\rm high} \sim 100$ are disfavored for explaining the quiescent emission from M87* and Sgr A* \citep[e.g.][]{Bower2003Interferometric,Marrone2007Unambiguous,eht2019m87iv,eht2021m87viii}, in part because such models have high densities and thus too little linear polarization due to Faraday depolarization.   This does not, however, rule out that {\em during flares}, models with $R=R_{\rm low} \sim R_{\rm high} \sim 100$ could be appropriate for describing the electron distribution in the near-horizon environment. 


In reality, plasma in near-horizon current sheets (i.e., at the base of a jet or magnetospheric region) and in the reconnection exhaust ejected into the disk and jet boundary, may consist of electron-positron pairs (as opposed to floored matter in GRMHD). This plasma is then heated by reconnection, up to temperatures similar to the jet's magnetization \citep{Ripperda2020,ripperda2022black} and limited by radiative cooling , instead of being limited by the numerically enforced $\sigma_{\rm floor}$ in GRMHD simulations. For highly magnetized plasma feeding the reconnection (e.g., $\sigma \geq 10^7$ for M87's jet, \citealt{hakobyan2022radiative}), the reconnection-accelerated particles heated to $T \sim \sigma$, are unlikely to emit much radiation at lower photon energies (e.g., in the mm or IR). Therefore, we argue that the high electron temperature models presented in this paper (e.g., $R = 1$) best capture the real physics of the reconnection heated plasma during magnetic flux eruptions.  These models predict a dimming of the millimeter emission during high-energy flares.
However, if the magnetization of the jet feeding the reconnection is much smaller $\sigma \ll 10^6$ \citep{2022PhRvL.129t5101C}, there may be enough non-thermal electrons emitting at submillimeter wavelengths to produce a flux comparable to the quiescent emission observed by \cite{eht2019m87i,eht2022sgri}. Models with $R=R_{\rm low} \sim R_{\rm high} \sim 100$ correspond in principle to this scenario of a less magnetized jet that feeds the reconnection layer, resulting in brightening of submillimeter wavelength emission during high-energy flares.

The properties of mm emission during magnetic flux eruptions depend most sensitively on the heating/acceleration of $\gamma \sim 100$ electrons, since those particles emit most of their synchrotron radiation in the mm.   As we have just argued, this 
is expected to be inefficient for reconnection in strongly magnetized plasmas \citep{Sironi2014Relativistic},
which would predict mm {dimming} coincident with high-energy flares.  
However, the interaction between the reconnecting current sheet and the bulk of the disk at somewhat larger radii is complex, could be sourced by less magnetized plasma, and could dominate the heating of plasma responsible for the mm emission.  It is also not at all clear that this interaction is well-modeled by existing GRMHD simulations (see \citealt{Galishnikova2022Collisionless} for a comparison of first principles general relativistic particle-in-cell (GRPIC) and GRMHD models of magnetic flux eruptions).

To understand better whether the mm emission during a flare can brighten, it will ultimately  be necessary to model the spatial and temporal dependence of the electron distribution, taking into account particle acceleration due to magnetic reconnection (and other processes) in the near-horizon environment.  The physics of particle heating and acceleration in the flare and quiescent states could also be significantly different (e.g., because the former is dominated by higher magnetization plasma than the latter).   This highlights a significant shortcoming of using a simple time-independent prescription $T_e(T_{\rm fluid})$ to model the emission and variability in systems like M87* and Sgr A*.  This is particularly true for MAD models that feature such physically distinct magnetic flux eruptions.

Daily bright and rapid flares have been observed from Sgr A* in X-ray, IR and mm wavelengths.   These show, however, different types of multi-band light curves in different flares, implying that they may be powered by different mechanisms.
For example, Figure 2 of \citet{fazio2018multiwavelength} shows that the mm \textit{brightens} simultaneously with the IR flare, consistent with the low electron temperature regime in this work.
On the other hand, Figure 1 of \citet{Yusef-Zadeh2010Occultation} shows that the mm \textit{dims} during the IR flare {(see also, \cite{2022ApJ...930L..19W}, for mm dimming during an X-ray flare}), consistent with the high electron temperature regime in this work.  Correlated changes in the image size and polarization as predicted in this paper would clarify whether the difference between these two types of flares is indeed the electron temperature the plasma is heated to during the flare.
Figure 24 of \citet{Yusef-Zadeh2009Simultaneous} and Figure 3 of \citet{Trap2011Concurrent} show a third type of phenomenology:  the mm flux does not change much during an IR flare, but increases later after a significant delay.  This is not captured by any electron temperature model in this work.   This further suggests that simple time-independent $T_e(T_{\rm fluid})$ prescriptions on top of ideal GRMHD simulations are not adequate to comprehensively explain the flare state observational signatures, which still requires better understanding of particle acceleration around accreting black holes.


\label{sec:discuss}

\section*{Acknowledgements}

We are grateful to Angelo Ricarte for helpful comments on our draft.
BR would like to thank Jordy Davelaar for useful discussions. EQ was supported in part by a Simons Investigator grant from the Simons Foundation.   This research was enabled by support provided by grant No. NSF PHY-1125915 along with a INCITE program award PHY129, using resources from the Oak Ridge Leadership Computing Facility, Summit, which is a US Department of Energy office of Science User Facility supported under contract DE-AC05- 00OR22725, as well as Calcul Quebec (http://www.calculquebec.ca) and Compute Canada (http://www.computecanada.ca).
The analysis presented in this article was performed in part on computational resources managed and supported by Princeton Research Computing, a consortium of groups including the Princeton Institute for Computational Science and Engineering (PICSciE) and the Office of Information Technology's High Performance Computing Center and Visualization Laboratory at Princeton University.
AP acknowledges support by NASA grant 80NSSC22K1054 and NSF grant PHY-2231698. This research was facilitated by Multimessenger Plasma Physics Center (MPPC), NSF grant PHY-2206607. The computational resources and services used in this work were partially provided by facilities supported by the Scientific Computing Core at the Flatiron Institute, a division of the Simons Foundation. This research is part of the Frontera computing project at the Texas Advanced Computing Center (LRAC-AST21006). Frontera is made possible by National Science Foundation award OAC-1818253. Support for this work was provided by NASA through the NASA Hubble Fellowship grant HST-HF2-51518.001-A awarded by the Space Telescope Science Institute, which is operated by the Association of Universities for Research in Astronomy, Incorporated, under NASA contract NAS5- 26555.

\section*{Data Availability}

The data underlying this paper will be shared on reasonable request
to the corresponding author.



\bibliographystyle{mnras}
\bibliography{example} 




\appendix

\section{Convergence Tests}

\begin{figure*}
	\includegraphics[width=15cm]{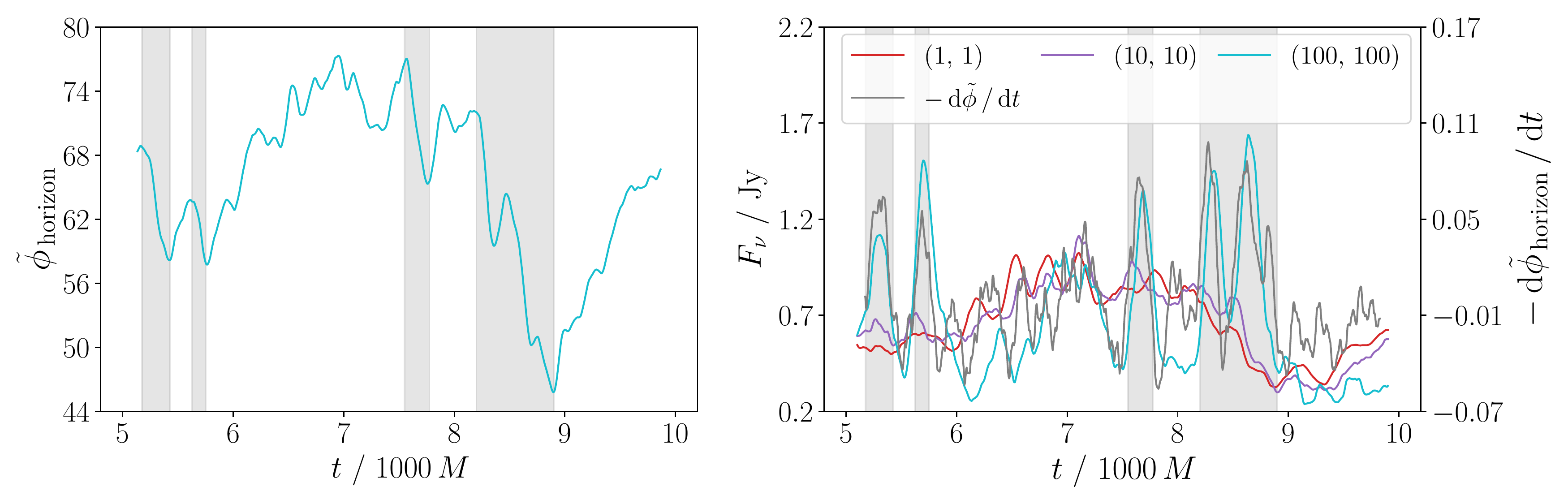}
    \caption{{\em Left:} the magnetic flux $\tilde{\phi}_{\rm horizon} \equiv \frac{1}{2} \int_0^{2\pi} \int_0^{\pi} \left| ^*F^{rt}\right|\sqrt{-g}{\rm d}\theta{\rm d}\phi$ on the black hole horizon, but for a separate \textit{standard} resolution GRMHD simulation.
    The grey bands indicate possible high energy flares in this simulation.
    {\em Right:} the evolution of 230 GHz flux with three ($R_{\rm high}$, $R_{\rm low}$) models, compared with $-\,{\rm d}\tilde{\phi}_{\rm \,horizon}\,/\,{\rm d}t$.
    Similar to the \textit{high} resolution GRMHD results, we find that 230 GHz flux with low $R$ electron temperature models is correlated with $\tilde{\phi}_{\rm horizon}$ and dims during high energy flares.
    On the other hand, 230 GHz flux with high $R$ electron temperature models is correlated with $-\,{\rm d}\tilde{\phi}_{\rm \,horizon}\,/\,{\rm d}t$ and brightens during high energy flares.}
    \label{fig:fluid-low}
\end{figure*}

\begin{figure*}
	\includegraphics[width=\textwidth]{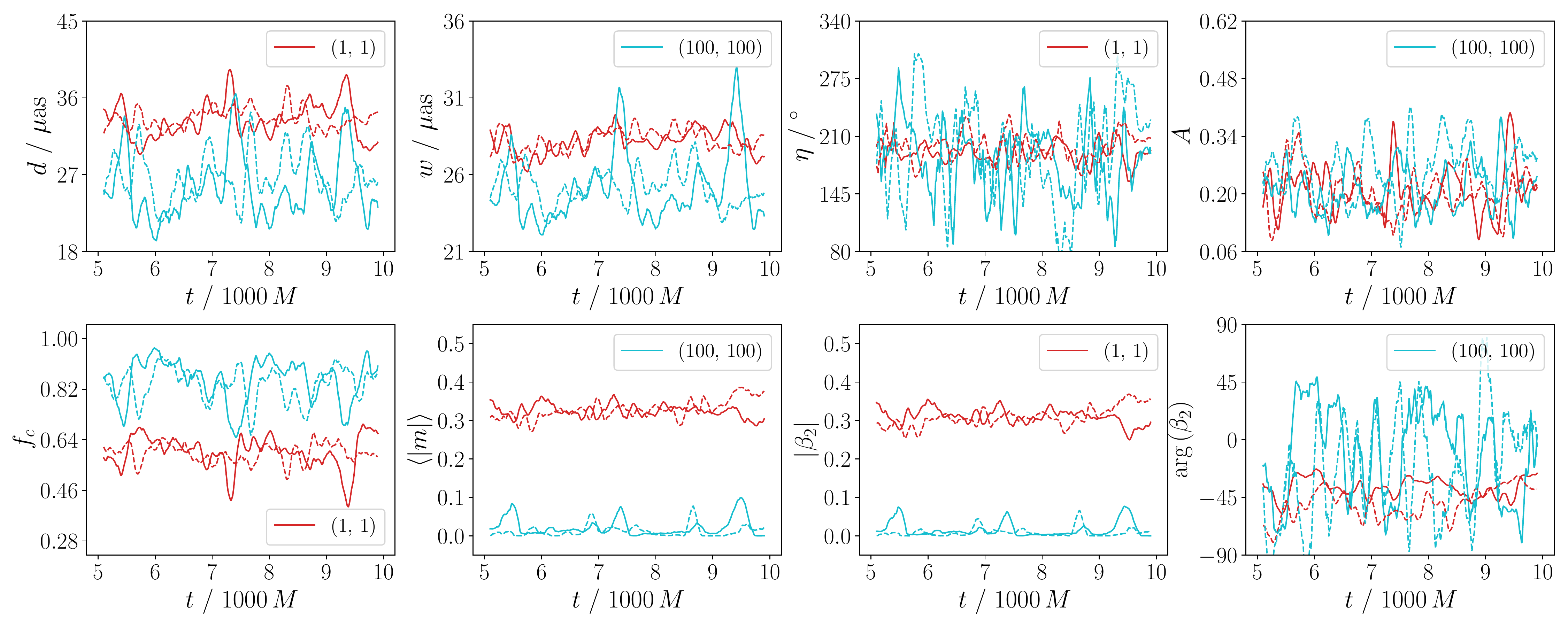}
    \caption{Comparison of the ring statistics in Figure \ref{fig:stats}, for the \textit{high} resolution (solid) vs \textit{standard} resolution (dashed) GRMHD simulations, with two ($R_{\rm high}$, $R_{\rm low}$) electron temperature models.
    Note that since here we have two separate GRMHD simulations, the ring statistics with the same $T_e$ model are not supposed to be exactly the same at each time, but they do look statistically similar.
    Therefore, the main results of this paper are also valid for lower resolution GRMHD simulations.}
    \label{fig:stats-03-01}
\end{figure*}

\label{sec:converge}

\subsection{GRMHD resolution convergence}

In this subsection, we check the sensitivity of our main results to GRMHD resolution, from two aspects.
First, we compare the ray tracing results with the \textit{high} resolution ($5376\times2306\times2306$) GRMHD simulation, to the results with a separate \textit{standard} resolution ($580\times288\times256$) GRMHD simulation, which has the same setup as the \textit{high} resolution simulation except that it was run at lower resolution.
As current EHT analyses \citep{eht2019m87v,eht2022sgri} are mostly done with simulations similar to the \textit{standard} resolution one here, it is important to check whether the our statistics have converged with respect to the GRMHD resoultion.
See Figures \ref{fig:fluid-low} and \ref{fig:stats-03-01}, we find no statistical difference between the ray tracing results with these two GRMHD simulations: they both brighten at 230 GHz during high energy flares with low $T_e$ models, and dim at 230 GHz during high energy flares with high $T_e$ models, while the other statistics are also similar.
Therefore, \textit{standard} resolution GRMHD simulations are probably adequate for current EHT analyses.

\begin{figure*}
	\includegraphics[width=\textwidth]{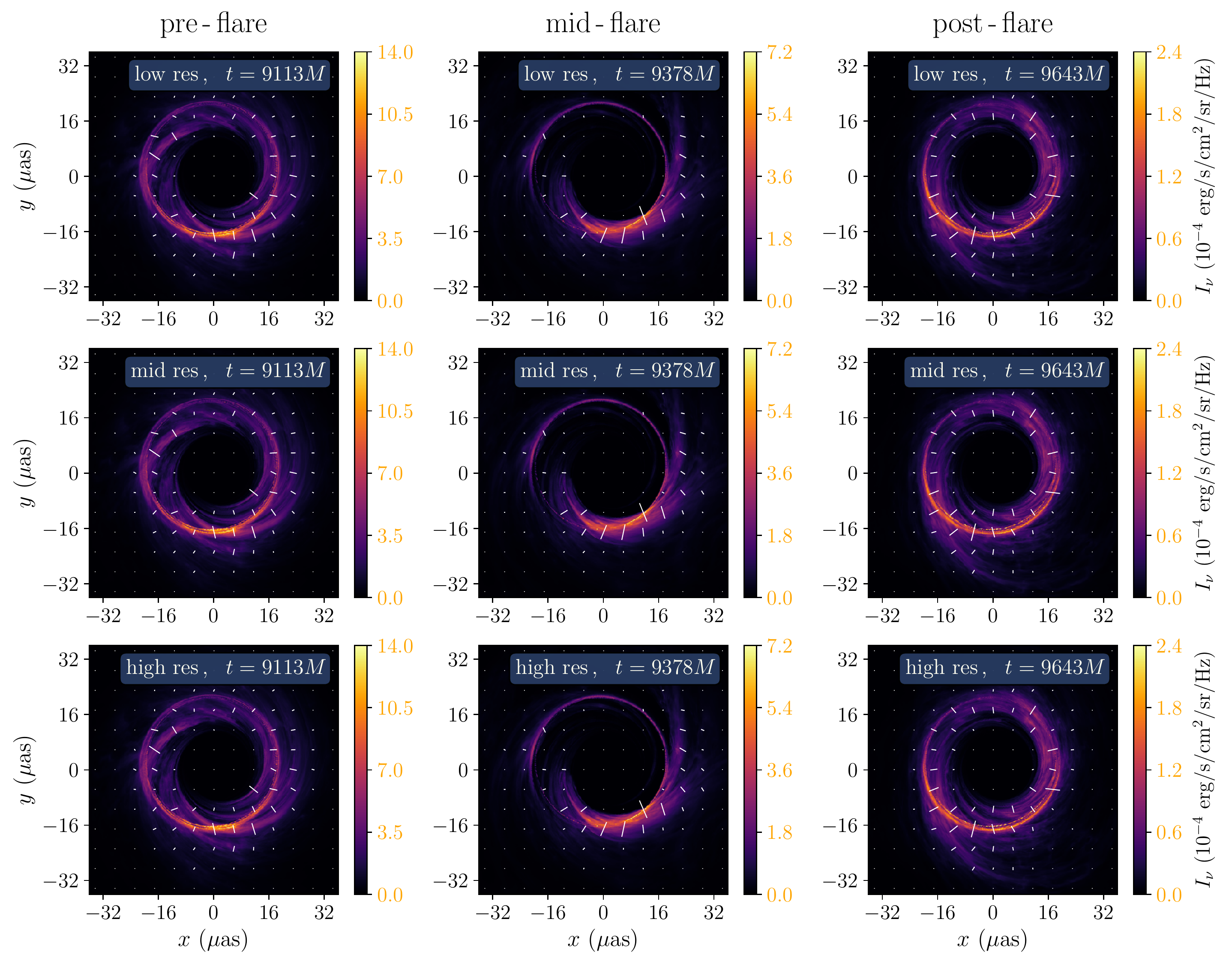}
	\caption{A convergence test of the intensity and polarization images with respect to the GRMHD resolution.
	For the "low res", "mid res" and "high res" setups, the GRMHD grid resolution in ray tracing is reduced by a factor of 8, 4 and 2, respectively, relative to the original GRMHD simulation resolution.
	We use $R=R_{\rm low}=R_{\rm high}=1$ for all the panels.
	The intensity and polarization maps are both very close between different GRMHD resolutions, so the "low res" setup is presumbly enough for theoretical modelling at EHT resolution.}
    \label{fig:polar-07-04}
\end{figure*}

\begin{figure*}
	\includegraphics[width=\textwidth]{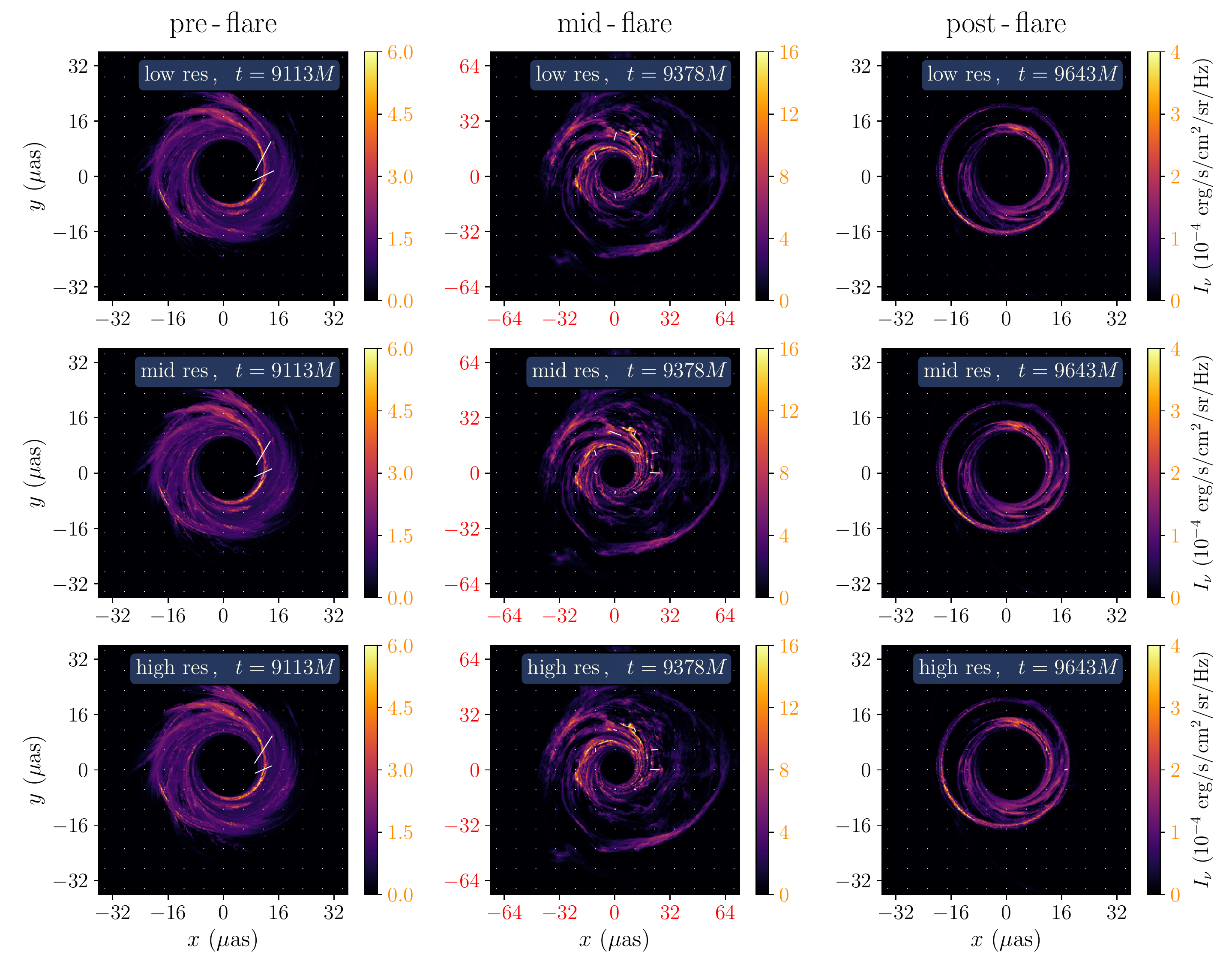}
	\caption{Similar to Figure \ref{fig:polar-07-04}, but using $R=R_{\rm high}=R_{\rm low}=100$ for the electron temperature.
	Here the difference in polarization maps is larger than Figure \ref{fig:polar-07-04}, as the emission comes from larger $T_{\rm fluid}$ regions where the direction of magnetic field is more disordered.
	However, such difference becomes less noticeable once the images are blurred at the EHT resolution (Figure \ref{fig:polar-07-06}).}
    \label{fig:polar-07-05}
\end{figure*}

Also, we note that during the ray tracing we reduce the GRMHD resolution by a factor of 4 (i.e. only keep one of the four successive points along each spatial dimension) to speed up the computation.
However, in the \textit{high} resolution GRMHD simulation, there are some small scale structures like plasmoids and X-points \citep[see Figure 1 of][]{ripperda2022black}, which may not be resolved in the reduced GRMHD data.
Here we test the convergence of ray tracing results with respect to GRMHD resolution, by comparing the images from "low res" (reduced by a factor of 8), "mid res" (reduced by a factor of 4, which is the default setting in this paper) and "high res" (reduced by a factor of 2) ray tracing runs.
Note that we only reduce the resolution of the same GRMHD data by different factors during ray tracing, and all the images are ray traced with a resolution of $512^2$ pixels.
The images with high temperature $R=1$ and low temperature $R=100$ models are shown in Figures \ref{fig:polar-07-04} and \ref{fig:polar-07-05}, respectively.
For $R=1$, both intensity and polarization maps are almost identical between the different downsampling of the high resolution GRMHD run, while for $R=100$ we do notice some moderate difference in the polarization maps, probably because the small scale plasmoids and X-points typically have higher $T_{\rm fluid}$ and are only visible at 230 GHz with lower electron temperature models.
However, when blurred with a 20$\,\mu$as Gaussian kernel, such differences diminish, since ray tracing with reduced GRMHD data can be regarded as a blurring in the GRMHD space, whose effective kernel size in the image space should be much smaller than 20$\,\mu$as as long as one only reduce the GRMHD data by a factor of $\lesssim\,$10.
Therefore, the default "mid res" settings should be adequate for EHT modelling and analysis.

\begin{figure*}
	\includegraphics[width=\textwidth]{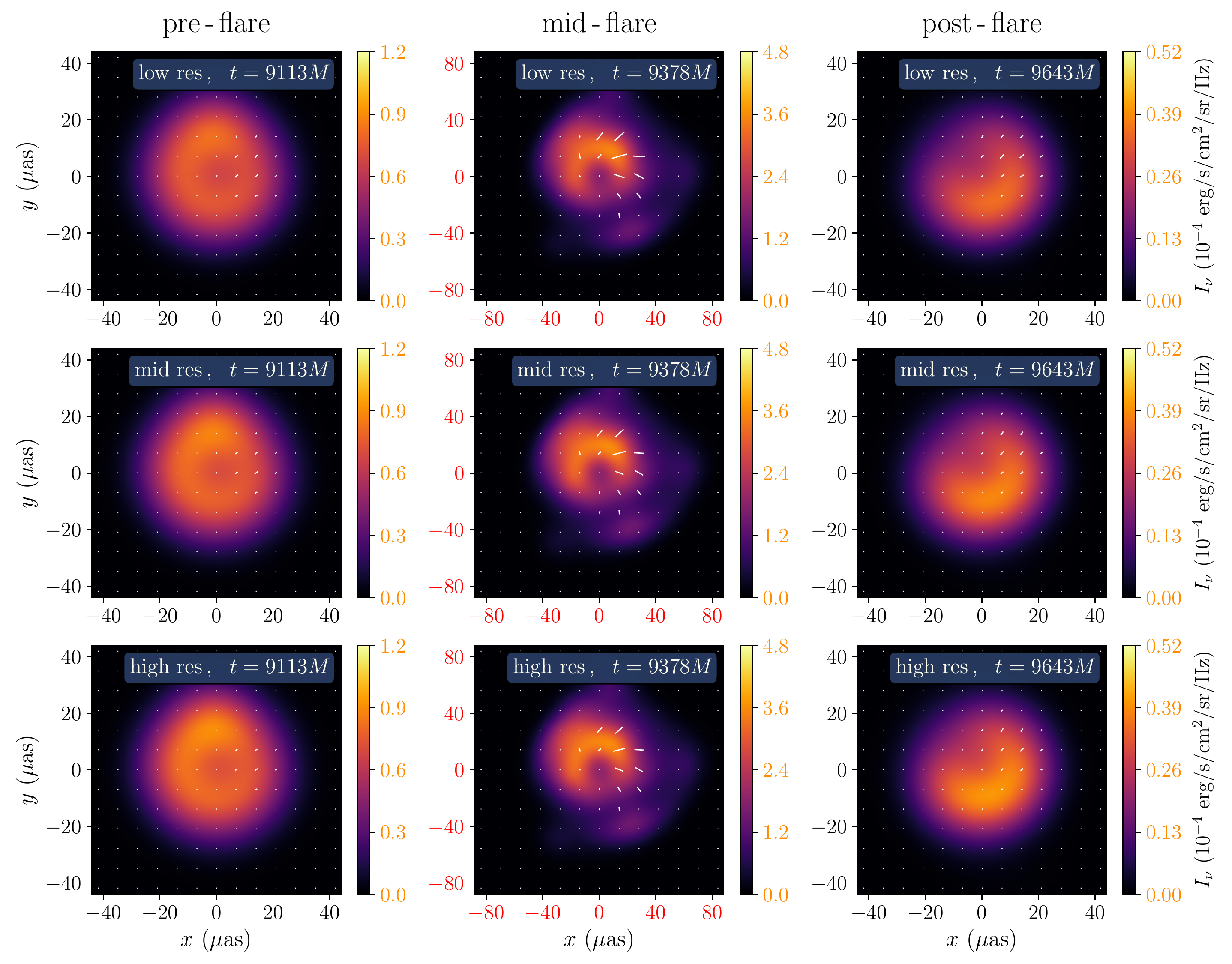}
	\caption{Similar to Figure \ref{fig:polar-07-05}, but blurred with a 20$\,\mu$as FWHM Gaussian kernel.
	The difference between different GRMHD resolution ray tracing diminishes after blurring.
	For the pre-flare and post-flare states, only a small arc-shaped region has non-negligible polarization, such that the pixel level polarization in the blurred images becomes much smaller than the mid-flare state.}
    \label{fig:polar-07-06}
\end{figure*}

\subsection{$\sigma_{\rm cut}$ convergence}

\label{sec:sigmacut}

\begin{figure*}
	\includegraphics[width=\textwidth]{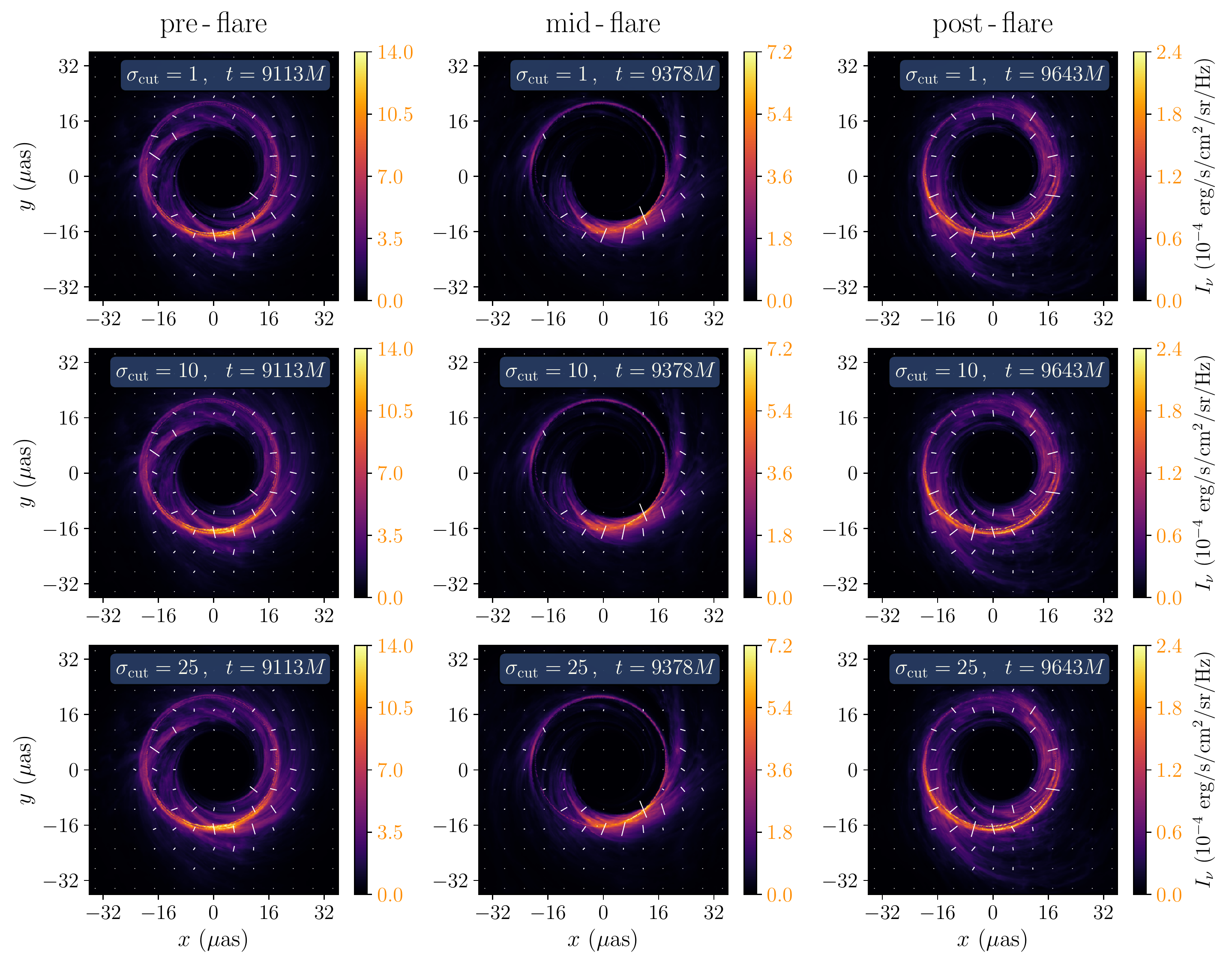}
	\caption{Comparison of images with different values of $\sigma_{\rm cut}$.
	We use $R=R_{\rm low}=R_{\rm high}=1$ for the electron temperature.
	Both intensity and polarization maps are quite similar between different $\sigma_{\rm cut}$'s.}
    \label{fig:polar-07-02}
\end{figure*}

\begin{figure*}
	\includegraphics[width=\textwidth]{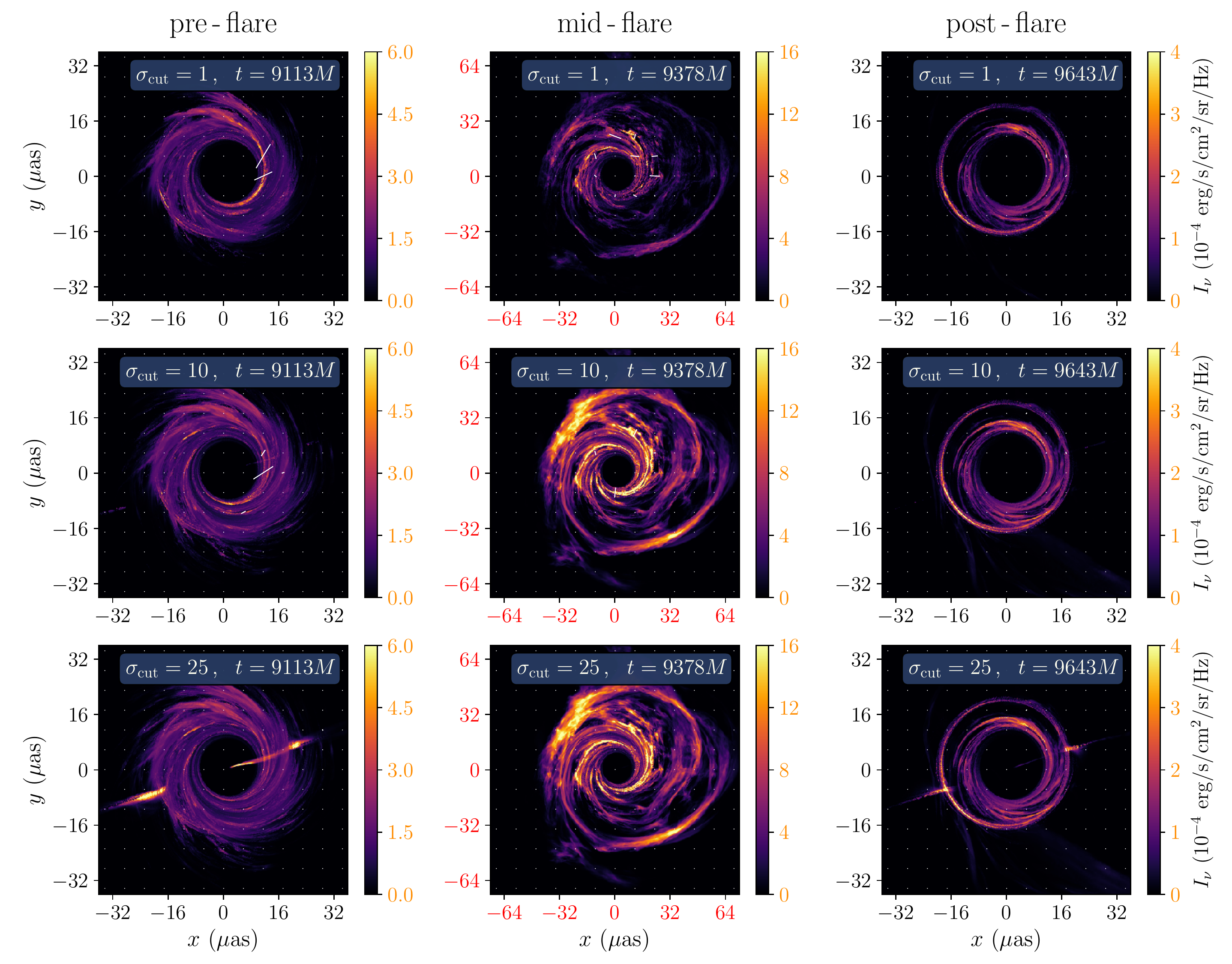}
	\caption{Similar to Figure \ref{fig:polar-07-02}, but using $R=R_{\rm high}=R_{\rm low}=100$ for the electron temperature.
	In this case we see bright, unphysical stripes with $\sigma_{\rm cut}=25$ which originate from the polar regions in the GRMHD simulation.
	While the flare state images are sensitive to the choice of $\sigma_{\rm cut}$, the main results of this paper based on a conservative choice of $\sigma_{\rm cut}$ should still be valid qualitatively, since using a larger $\sigma_{\rm cut}$ will only make mid-flare images even brighter than the quiescent states.}
    \label{fig:polar-07-03}
\end{figure*}

In this paper, we apply $\sigma_{\rm cut}=1$ to the 230 GHz ray tracing calculations in Section \ref{sec:lightcurve}-\ref{sec:image}, and $\sigma_{\rm cut}=10$ to the spectra calculations in Section \ref{sec:multi}.
Here we check the convergence with respect to $\sigma_{\rm cut}$ by comparing the ray tracing images using $\sigma_{\rm cut}=1$, 10 and 25, respectively.
For $R=1$, we do not see any noticeable difference between the images with the three different $\sigma_{\rm cut}$'s (Figure \ref{fig:polar-07-02}), indicating that $\sigma_{\rm cut}=1$ should be a self-consistent choice for such "standard" electron temperature models.
On the other hand, for $R=100$, we find unphysical (i.e., governed by the GRMHD $\sigma_{\rm floor}$ and/or density/pressure injection in polar regions) bright stripes in the quiescent state images with $\sigma_{\rm cut}=25$, which are also minimally noticeable in $\sigma_{\rm cut}=10$ images but are gone with $\sigma_{\rm cut}=1$ (Figure \ref{fig:polar-07-03}).
In the mid-flare state, the image with $\sigma_{\rm cut}\gtrsim10$ is much brighter than $\sigma_{\rm cut}=1$, since $\sigma_{\rm cut}$ is likely activated more frequently during the flares.
As shown in the right panel of Figure \ref{fig:TL}, higher $T_{\rm fluid}$ plasma is more likely to have $\sigma\sim\sigma_{\rm cut}$, which explains why at 230 GHz $\sigma_{\rm cut}$ is more likely to affect the images with lower electron temperature models.

Unfortunately, realistic SMBH accretion flows may have much larger $\sigma \gg \sigma_{\rm floor}$, which is far beyond the capability of GRMHD simulations.
Therefore, the results presented in this paper (especially those using low $T_e$ models and/or at high frequencies) should be regarded as a lower bound of the actual emission, as we make a conservative choice of $\sigma_{\rm cut}$ to eliminate the unphysical emissions, which however may also remove part of the signal that is indeed physical.
It is yet not clear whether the unmodeled high $\sigma$ region would significantly change the black hole observational signatures, which we leave for future research.


\bsp	
\label{lastpage}
\end{document}